\documentclass[fleqn,usenatbib]{mnras}

\usepackage{newtxtext,newtxmath}

\usepackage[T1]{fontenc}

\DeclareRobustCommand{\VAN}[3]{#2}
\let\VANthebibliography\thebibliography
\def\thebibliography{\DeclareRobustCommand{\VAN}[3]{##3}\VANthebibliography}

\usepackage{graphicx}	
\usepackage{amsmath}	

\title[Spectral-timing study of GRS 1915+105]{Comprehensive X-ray Spectral-timing Analysis of GRS 1915+105 Based on Insight-HXMT Observations}

\author[X. Chen et al.]{
Xiao Chen,$^{1,2}$
Weiping Liu$^{1,3}$\thanks{E-mail: liuwp@sustech.edu.cn} and
Wei Wang$^{2}$\thanks{E-mail: wangwei2017@whu.edu.cn}
\\
$^{1}$Department of Physics, College of Science, Southern University of Science and Technology, Shenzhen 518055, China\\
$^{2}$Department of Astronomy, School of Physics and Technology, Wuhan University, Wuhan 430072, China\\
$^{3}$China Institute of Atomic Energy, P.O. Box 275(11), Beijing 102413, China
}

\date{Accepted XXX. Received YYY; in original form ZZZ}

\pubyear{2015}

\begin{document}
\label{firstpage}
\pagerange{\pageref{firstpage}--\pageref{lastpage}}
\maketitle

\begin{abstract}
GRS 1915+105 has been well studied since its discovery, and is well-known for its complex light curve variability. Using the full currently available Insight-HXMT dataset from July 2017 to June 2023, we make a comprehensive spectral-timing analysis of this source and report four main findings. First, we uncover a QPO frequency rising branch between MJD 58206 and 58230, where the centroid frequency increases from $\sim$2~Hz to $\sim$6~Hz, consistent with a spectral state transition from the hard to intermediate state. This rising branch completes the full QPO frequency evolution cycle when combined with the subsequent frequency decay phase, and had been missed in prior NICER and Insight-HXMT studies. Second, we identify a previously unreported Flare 3 during the obscured state, which shows distinct spectral and timing properties compared to the earlier flares. Third, we detect sub-Hz QPOs (<1 Hz) in all three flares, specifically at $\sim$0.01 Hz in Flare 1 and $\sim$0.2 Hz in both Flares 2 and 3. In particular, the weak $\sim$0.2 Hz signals observed in Flare 3 indicate ongoing coronal activity despite strong obscuration. Finally, a comparison between QPOs above and below 1~Hz suggests distinct origins, with the former likely arising from Lense–Thirring precession of the inner hot flow and the latter from magnetic perturbations driving a failed disk wind. These findings offer new insights into the unique accretion geometry and variability behaviors of GRS 1915+105.

\end{abstract}




\section{Introduction}
\label{sec:intro}

Black hole X-ray binaries (BHXRBs) are systems in which a black hole accretes matter from a companion star, producing significant X-ray emission. Most BHXRBs are transients and have relatively short outbursts. They are sufficiently bright in X-ray for only a few weeks to months \citep{Remillard1999ApJ...522..397R,Homan2001ApJS..132..377H}. A typical outburst evolve though different spectral states -— including the hard state, intermediate states, and soft state -- and traces a q-shaped pattern in the hardness-intensity diagram \citep{Homan2001ApJS..132..377H,Fender2006MNRAS.369..603F}. The hard state is dominated by a power-law component and often accompanied by low-frequency quasi-periodic oscillations (QPOs), while the soft state is dominated by thermal disk emission. Among low-frequency QPOs, type-C QPOs are the most common type, with centroid frequencies typically in the 1--10~Hz range. Type-B QPOs are weaker, usually centered around 5--6~Hz, and are believed to be associated with the soft intermediate state \citep{Belloni2016ASSL..440...61B}. Their power density spectra (PDS) often show weak red noise below 0.1~Hz \citep{Ingram2019NewAR..8501524I}.

GRS~1915+105, a prototypical microquasar, was discovered in 1992 \citep{Castro-Tirado1994ApJS...92..469C}. Since then, it has exhibited remarkable X-ray variability, QPOs, superluminal radio jets, and highly ionized winds \citep{Mirabel1994Natur.371...46M,Morgan1997ApJ...482..993M,Belloni2000A&A...355..271B,Lee2002ApJ...567.1102L,Ueda2009ApJ...695..888U,Harikesh2025MNRAS.tmp..910H}. It is a low-mass BHXRB system hosting a K-type giant donor star, with a measured black hole mass of $12.4^{+2.0}_{-1.8}$ solar masses and a distance of $8.6^{+2.0}_{-1.6}$~kpc \citep{Greiner2001A&A...373L..37G,Reid2014ApJ...796....2R}. GRS 1915+105 has the largest accretion disk and the longest orbital period among low-mass BHXRBs -- 33.9 days \citep{Steeghs2013ApJ...768..185S}. Unlike classical BHXRBs \citep{Belloni2016ASSL..440...61B}, it transitions between three spectral states labeled A, B and C \citep{Belloni2000A&A...355..271B}. States A and B are soft, with limited X-ray variability and significant disk contribution, while state C is hard, with strong X-ray variability, little or no disk contribution, and steady radio jets \citep{Rushton2010A&A...524A..29R}.

\begin{figure*}
    \centering
    \includegraphics[width=\textwidth]{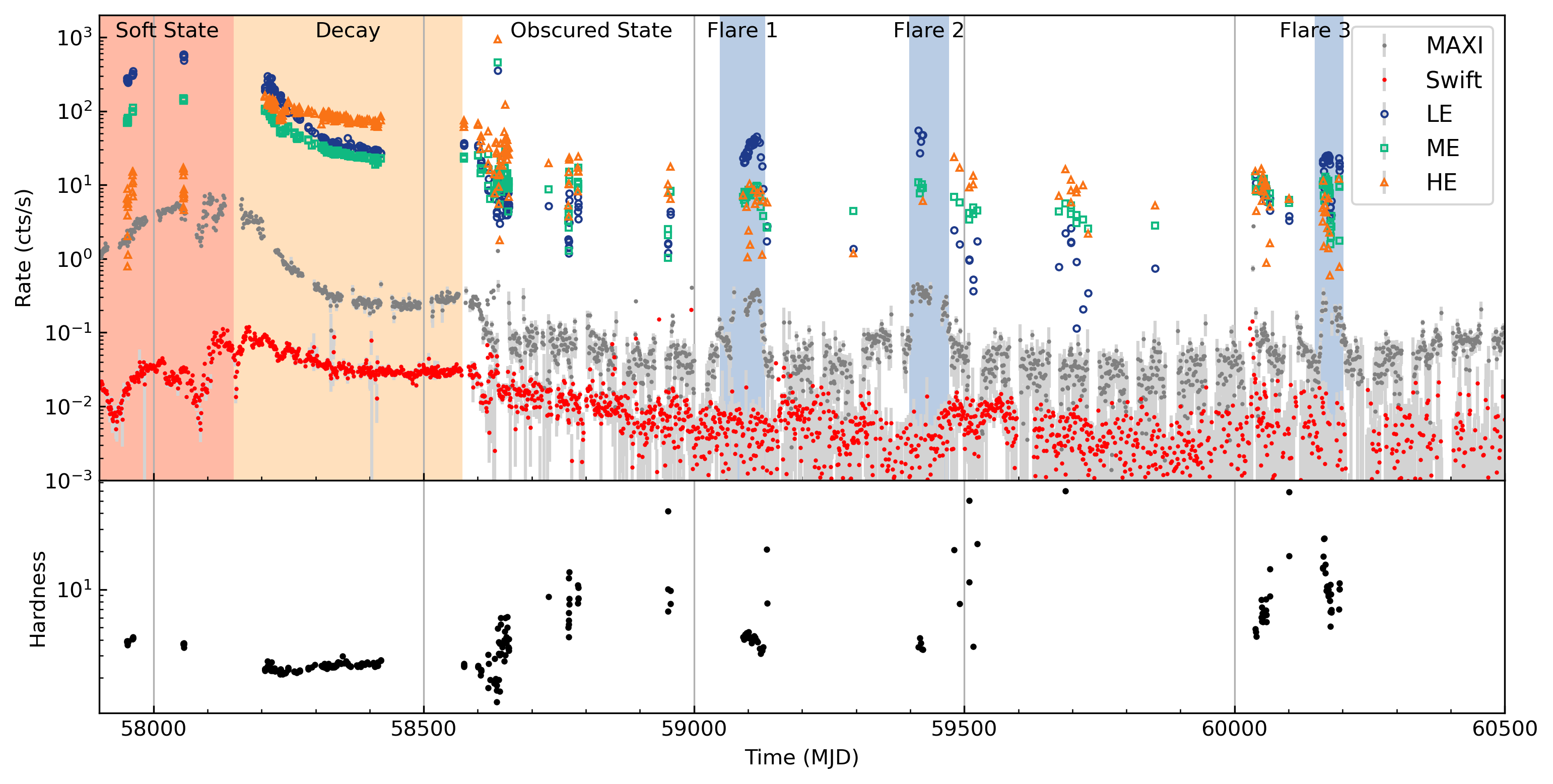}
    \caption{{\bf Top panel}: Light curves from LE (dark blue), ME (emerald green), HE (orange), MAXI (grey), and Swift/BAT (red) are shown. All data points are shown with lightgrey error bars. The MAXI data correspond to the 2--20~keV energy range with units of ph~s$^{-1}$~cm$^{-2}$, and the BAT data cover the 15--50~keV range with units of count~s$^{-1}$~cm$^{-2}$. However, these units are not explicitly shown in the plot to avoid clutter. The y-axis label only indicates the units for the three Insight-HXMT detectors. Different states are marked with different background colors, with Soft Peach Pink represent the Active State (or Soft State), Pale Peach represent the Decay Phase, while Soft Blue represent the Flares. The Obscured Faint State are displayed with no (or white) background. {\bf Bottom panel}: Hardness ratio Evolution. The hardness ratio is calculated based on 3--10 keV / 1--3 keV rates in LE.}
    \label{fig:evolve_lc}
\end{figure*}

GRS~1915+105 remained bright in X-ray for nearly 26~years. Its outburst ended in 2018 and entered a so-called "obscured" state in 2019, providing a rare opportunity to study black hole accretion under different physical conditions (also see Figure \ref{fig:evolve_lc}). Although the X-ray flux decreased by over an order of magnitude compared to its historical high, the source is still believed to be active, as multiwavelength flaring has been reported in radio, infrared and X-rays \citep{Koljonen2019ATel12839....1K,Trushkin2019ATel12855....1T,Motta2019ATel12773....1M,Motta2021MNRAS.503..152M,Gandhi2025MNRAS.537.1385G}. During this phase, strong partial covering absorption has been detected \citep{Miller2019ATel12771....1M,Balakrishnan2021ApJ...909...41B,Koljonen2020A&A...639A..13K,Miller2020ApJ...904...30M}, which is a rare feature among BHXRBs. The obscuration is thought to cause the X-ray dimming, though its geometry and physical nature remain debated \citep{Negoro2018ATel11828....1N,Miller2020ApJ...904...30M,Balakrishnan2021ApJ...909...41B,Motta2021MNRAS.503..152M,Sanchez-Sierras2023A&A...680L..16S,Neilsen2020ApJ...902..152N}. In other binary systems like V404 Cyg and SS 433, similar obscuration has been linked to super-Eddington accretion or jet-driven outflows \citep{Spencer1979Natur.282..483S,Miller-Jones2019Natur.569..374M}.

QPOs in GRS 1915+105 exhibit rich variability across different accretion states and energy bands, providing key insights into the dynamics of the inner accretion flow. During the decay phase, low-frequency X-ray QPOs around 1-6 Hz were still detected with NICER \citep{Koljonen2021A&A...647A.173K}, but disappeared after the transition into the obscured state. However, X-ray QPOs at $\sim$ 0.02~Hz and 0.2~Hz were detected during the re-brightening events near MJD~59100 and MJD~59400, respectively \citep{Athulya2023MNRAS.525..489A,Kong2024A&A...686A.211K}. Meanwhile, radio QPOs at $\sim 5$Hz, $\sim 0.06$ and $0.03$ Hz during the obscured state are also reported with the Five-hundred-meter Aperture Spherical radio Telescope (FAST) in the GHz bands \citep{tian2023,wang2025}. The physical origin and energy dependence of these low-frequency QPOs, particularly those seen during the obscured state, remain unclear.

The Hard X-ray Modulation Telescope (Insight-HXMT) has provided extensive broadband observations of GRS 1915+105, yielding valuable insights into its complex accretion and ejection behaviors. Studies based on HXMT have revealed flares, QPO evolution, ionized disk winds, new variability patterns, and long-term variability properties \citep{Kong2021ApJ...906L...2K,Liu2021ApJ...909...63L,Liu2022ApJ...933..122L,Shi2023MNRAS.525.1431S,Zhou2025A&A...694A.104Z}. In this work, we conduct a comprehensive analysis of all publicly available HXMT data from MJD~57951 (July 17, 2017) to MJD~60193 (September 6, 2023). Section 2 describes the observations and data reduction. In Section 3, we present our timing and spectral results, including possible type-B QPOs, the increase in QPO frequency at the beginning of the decay phase, and a third weak flare -- all of which have not been reported before. We also apply wavelet analysis to investigate transient and mHz/sub-Hz QPOs during the obscured state, which are difficult to study using traditional PDS or even dynamical PDS. In Section 4, we discuss the QPO evolution and implications of these findings, particularly the new insights provided by these previously unreported features. Conclusions are summarized in Section 5.

\section{Observations and Data Reduction}
\label{sec:obs}

Insight-HXMT, China's first X-ray satellite, is equipped with three detectors onboard, i.e. the Low Energy (LE) detector \citep{Chen2020SCPMA..6349505C}, the Medium Energy (ME) detector \citep{Cao2020SCPMA..6349504C}, and the High energy (HE) detector \citep{Liu2020SCPMA..6349503L}, which cover broad energy ranges of 1--15~keV, 5--30~keV and 25--250~keV, respectively. The HE comprises 18 NaI/CsI detectors with a total geometrical area of approximately 5100~cm$^2$ and a typical field of view (FoV) of 1.1$^\circ$~$\times$~5.7$^\circ$. The ME is composed of 1728 Si-PIN detectors, providing a collecting area of 952~cm$^2$ and a FoV of 1$^\circ$~$\times$~4$^\circ$. The LE uses swept charge devices (SCDs), offering a total collecting area of 384~cm$^2$ and a FoV of 1.6$^\circ$~$\times$~6$^\circ$ \citep{Zhang2020SCPMA..6349502Z}.

As of April 2025, a total of 174 public HXMT observations of GRS~1915+105 are available, with a cumulative exposure of 5466 ks. All HXMT data were processed using HXMTDAS v2.05 and calibration database v2.06. Good time interval (GTI) selection criteria included: elevation angle $>10^\circ$, geomagnetic cutoff rigidity (COR) $>8$ GeV, off-axis angle $<0.04^\circ$, and the exclusion of intervals within 300 seconds of South Atlantic Anomaly (SAA) passages. Light curves were generated in the 1--10~keV, 10--35~keV, and 28-100~keV bands for LE, ME, and HE, respectively, with a time resolution of 0.0078125 s. Instrumental backgrounds were estimated by LEBKGMAP \citep{Liao2020JHEAp..27...24L}, MEBKGMAP \citep{Guo2020JHEAp..27...44G}, and HEBKGMAP \citep{Liao2020JHEAp..27...14L} provided in HXMTDAS v2.05 for LE, ME and HE, respectively, and were removed in light curves and during spectral fittings.

For timing analysis, PDSs were calculated for LE, ME and HE using 64-second segments with \texttt{Powspec}. Fractional rms normalization \citep{Miyamoto1991ApJ...383..784M} was performed and Poisson noise was subtracted. QPOs were modeled with multiple Lorentzians in \texttt{Xspec v12.14.1} \citep{Arnaud1996ASPC..101...17A} then. The QPO significance was estimated as the ratio between the normalization of the QPO Lorentzian and its 1 $\sigma$ uncertainty \citep{Belloni2012MNRAS.426.1701B}.

Wavelet analysis is a time-frequency decomposition technique that enables the detection and characterization of non-stationary and transient signals. Unlike the Fourier transform, which assumes signal is stationary in time and provides only global frequency information, wavelet transforms retain both time and frequency resolution, making them particularly well-suited for studying the evolving and short-lived variability commonly observed in accreting systems.Therefore, it is used in this work to analyze light curves in the obscured state, which may contains short-lived or transient QPO signals. It works by convolving the original signal with a set of wavelet functions that are generated by scaling and translating a chosen mother wavelet. This procedure produces a two-dimensional time-frequency result, known as the wavelet power spectrum (WPS), which reveals the evolution of signal power across different time and frequency scales. In this work, we employ the Morlet wavelet as the mother wavelet. It offers an good balance between time and frequency resolution, making it especially suitable for detecting QPOs with variable frequencies and duration that may be missed in conventional PDS methods. A more comprehensive introduction and step-by-step guide can be refereed in \cite{Torrence1998BAMS...79...61T} and \cite{Chen2022MNRAS.513.4875C}.

To assess the significance of features in the WPS, we calculate the 95\% confidence level assuming a red-noise background spectrum using the lag-1 autoregressive \citep{Torrence1998BAMS...79...61T}. Only features above this confidence level are considered to be significant. We also account for the edge effects in the wavelet transform by marking the cone of influence (COI), within which the results may be affected by boundary effects and should be treated with caution.

For spectral analysis, the energy ranges of 2--9 keV for LE, 10--20 keV for ME, and 30--50 keV for HE were included. As indicated in previous HXMT studies, the HE count rate is generally low and is heavily affected by high instrumental background during the soft state of GRS~1915+105 \citep{Liu2021ApJ...909...63L, Liu2022ApJ...933..122L, Zhou2025A&A...694A.104Z}. A similar situation is observed in our data during flares and parts of the obscured state. Even when background conditions are relatively favorable, the effective HE coverage in the obscured state remains narrow, typically constrained to below 50~keV. In the decay phase, HE signals extend up to $\sim$100 keV \citep{ Zhou2025A&A...694A.104Z}, but to ensure consistency across all states, we limited the HE band to 30–50 keV in our spectral analysis. The uncertainties of the spectral fitting parameters were estimated using Monte Carlo Markov Chain within \texttt{XSPEC}.

\section{Results}
\label{sec:results}

\subsection{Timing analysis}

\subsubsection{Light curve variability}
Figure~\ref{fig:evolve_lc} displays the long-term light curves of GRS~1915+105 monitored by Insight-HXMT from 2017 to 2023, alongside data from the Monitor of All-sky X-ray Image \citep[MAXI, shown as grey points;][]{Matsuoka2009PASJ...61..999M} in the energy range of 2--20~keV, and from \textit{Swift}/BAT \citep[red points;][]{Gehrels2004ApJ...611.1005G} in the 15--50~keV range. The MAXI light curve shows that GRS 1915+105 remained in its normal active stage prior to 2018. It then entered a decay phase, initially characterized by a rapid exponential decline, followed by a more gradual linear decay \citep{Koljonen2021A&A...647A.173K}. After $\sim$ MJD~58600, the source became significantly fainter, but exhibited several short or long lasting flares or re-brightenings \citep{Kong2021ApJ...906L...2K, Neilsen2020ApJ...902..152N, Zhou2025A&A...694A.104Z}. Two major flares have been reported in the former research \citep{Athulya2023MNRAS.525..489A}, but a third possible flare around MJD 60150 is apparent in our plot and has not been studied before. We refer to this newly identified event as Flare 3 in this work. Weak mini-burst(s) may also precede this flare. The LE and ME count rates show a consistent trend with the MAXI light curve, but the HE light curve exhibits subtle differences during the three flares. In some cases, it does not show a increase in brightness, or appears to lag behind, as seen following Flare 2. A similar pattern can be also observed in the \textit{Swift}/BAT data. The hardness ratio of 3--10 keV to 1--3 keV decreases after the soft state, but gradually increases during the decay phase, showing a more significant increase during the obscured state. Notably, the hardness ratio remains relatively low during the first two flares. However, during the third flare, the hardness ratio starts from a much higher value and then drops significantly over time, indicating that Flare 3 exhibits distinct light curve behaviors compared to the earlier events.

\begin{figure}
    \begin{minipage}{0.48\textwidth}
		\includegraphics[width=1\textwidth]{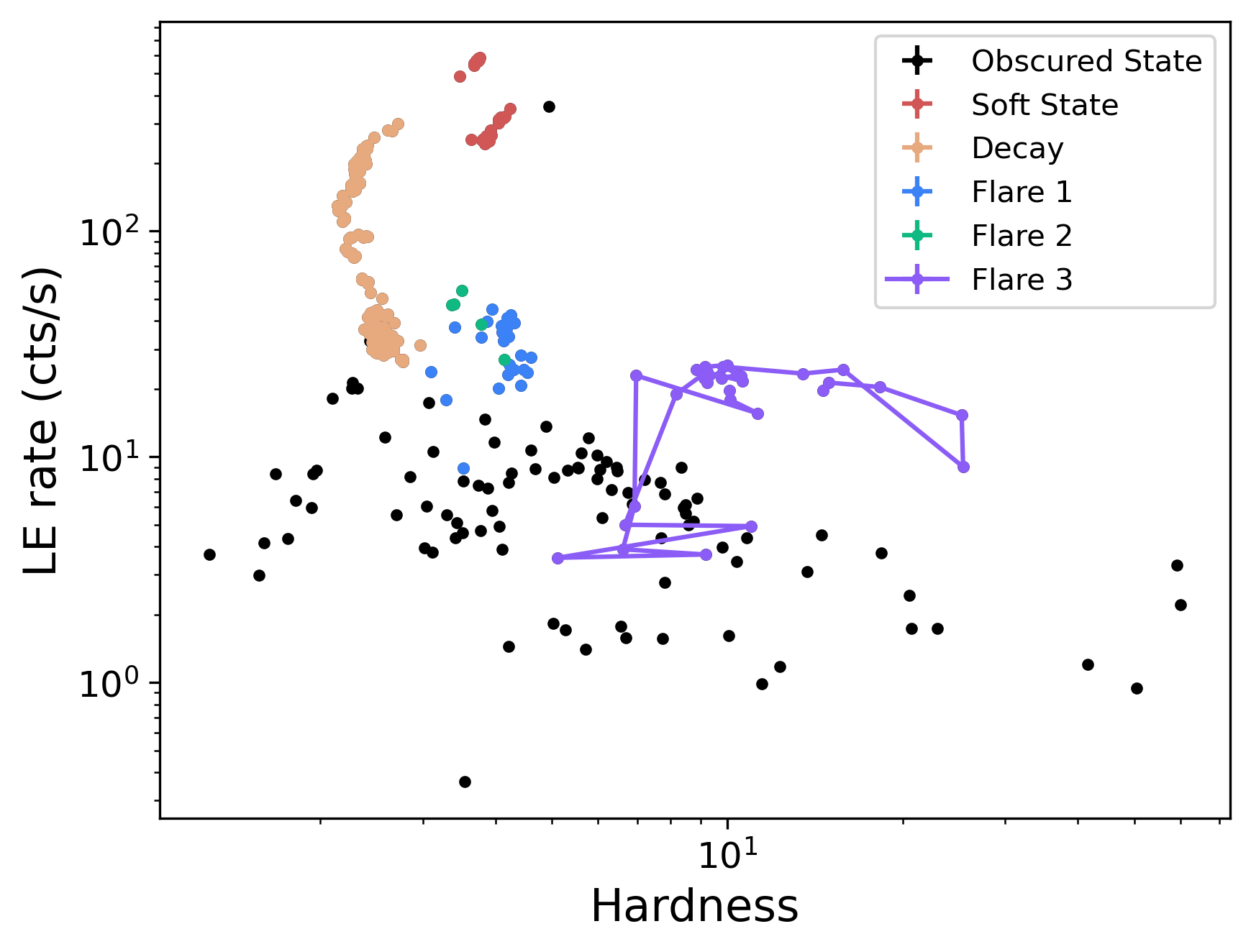}
	\end{minipage}
	\begin{minipage}{0.48\textwidth}
		\includegraphics[width=1\textwidth]{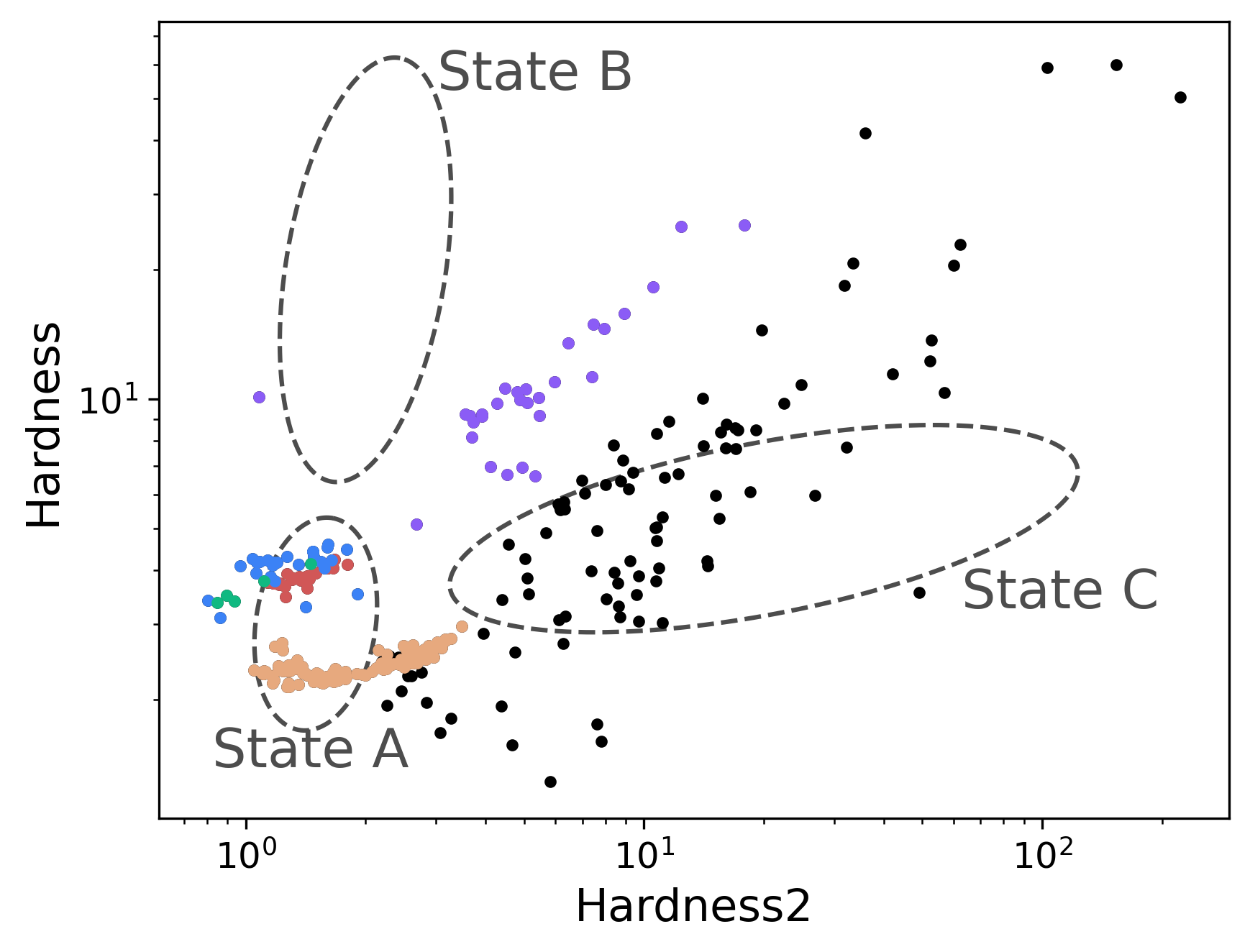}
	\end{minipage}
    \caption{{\bf Top panel}: Relation between the hardness ratio (3--10~keV to 1--3~keV) and LE (1--10~keV) count rate. Data from the Soft State, Decay Phase, Flare 1, Flare 2 and Flare 3 are shown in red, light tan, blue, green and purple, respectively, with the Flare 3 points connected in chronological order. The black points represent the obscured faint states. All data points are shown with error bars, although they are too small to be clearly visible in the plot. {\bf Bottom panel}: The relation between two X-ray colors. Hardness is defined as in the top panel, while Hardness2 is defined as the ratio of 10--35~keV to 1--3~keV count rate. The gray dashed ellipses indicate the approximate regions of the three states defined by \citet{Belloni2000A&A...355..271B}.}
    \label{fig:hr}
\end{figure}

To further investigate the light curve properties, we plot the relations between different hardness ratios and between hardness ratio and LE count rate in Figure~\ref{fig:hr}. As shown in the top panel, the source starts in the upper left corner, and gradually decrease in LE rate. While Flare 1 and 2 exhibit similar behavior in the plot, Flare 3 is quite different. Although the LE rate during Flare 3 is comparable to (or slightly lower than) that of the first two flares, its hardness ratio spans a much broader range. Some points during Flare 3 exhibit much lower count rates, forming a trajectory that resembles a q-shape in the plot. After connecting the points of Flare 3 chronologically, we find that it evolves from the upper right to the upper left, and then to the lower left. However, it subsequently returns to the upper left, indicating an increase in luminosity. Such behavior is more complex than that of a typical BHXRB. To investigate this further, we plot the hardness ratio of 3--10~keV to 1--3~keV (Hardness) against the ratio of 10--35~keV to 1--3~keV (Hardness2), following a method similar to \cite{Belloni2000A&A...355..271B}. In their scheme, State B is characterized by high count rate and high Hardness, State C by low count rate and low Hardness with variable Hardness2, and State A by low count rate, low Hardness, and low Hardness2. For clarity, we overplot the approximate regions of States A, B, C as defined by \cite{Belloni2000A&A...355..271B}. Based on their definitions, Flare 3 appears to lie between State B, the outburst state, and State C, the quiescent state, while Flare 1 and 2 are more consistent with State A. We note, however, that their state classification was derived from fast variability analysis of light curves with 1~s bins, whereas in our case each data point represents a full observation, which may naturally lead to some differences.

\subsubsection{QPOs above 1~Hz}

\begin{figure}
    \centering
    \includegraphics[width=\columnwidth]{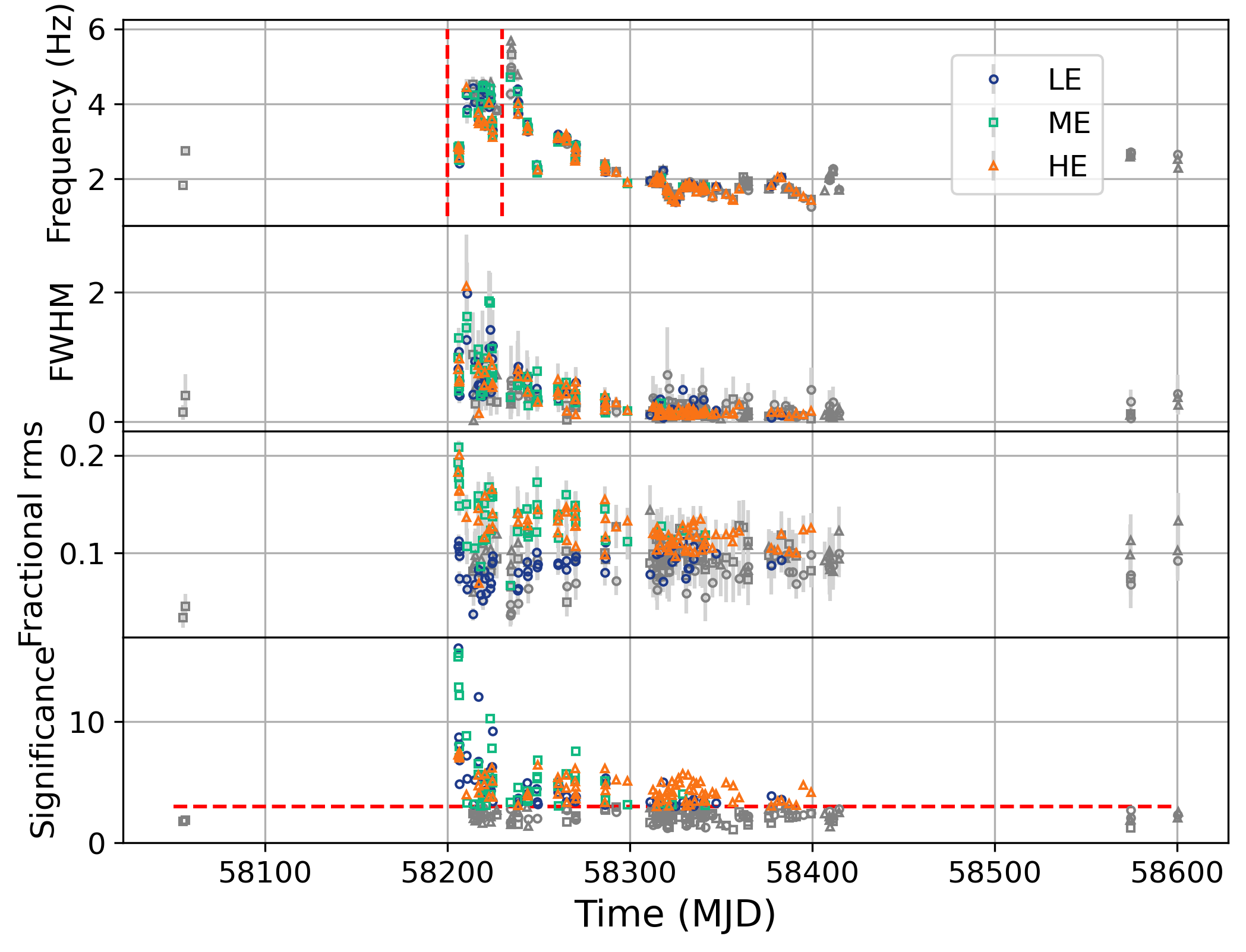}
    \caption{From top to bottom, the panels show the evolution of QPO centroid frequency above 1 Hz, FWHM, fractional rms, and significance over time. Data points with significance $\geq$3 are plotted in blue circles (LE), green squares (ME), and orange triangles (HE), while those with significance $<$3 are shown in grey. All data points are shown with light grey error bars. In the top panel, the region enclosed by the red dashed lines highlights the rising phase of the QPO centroid frequency. While in the bottom panel, the horizontal red dashed line marks the 3 $\sigma$ threshold.}
    \label{fig:evolve_QPO}
\end{figure}

\begin{figure*}
    \centering
    \includegraphics[width=\textwidth]{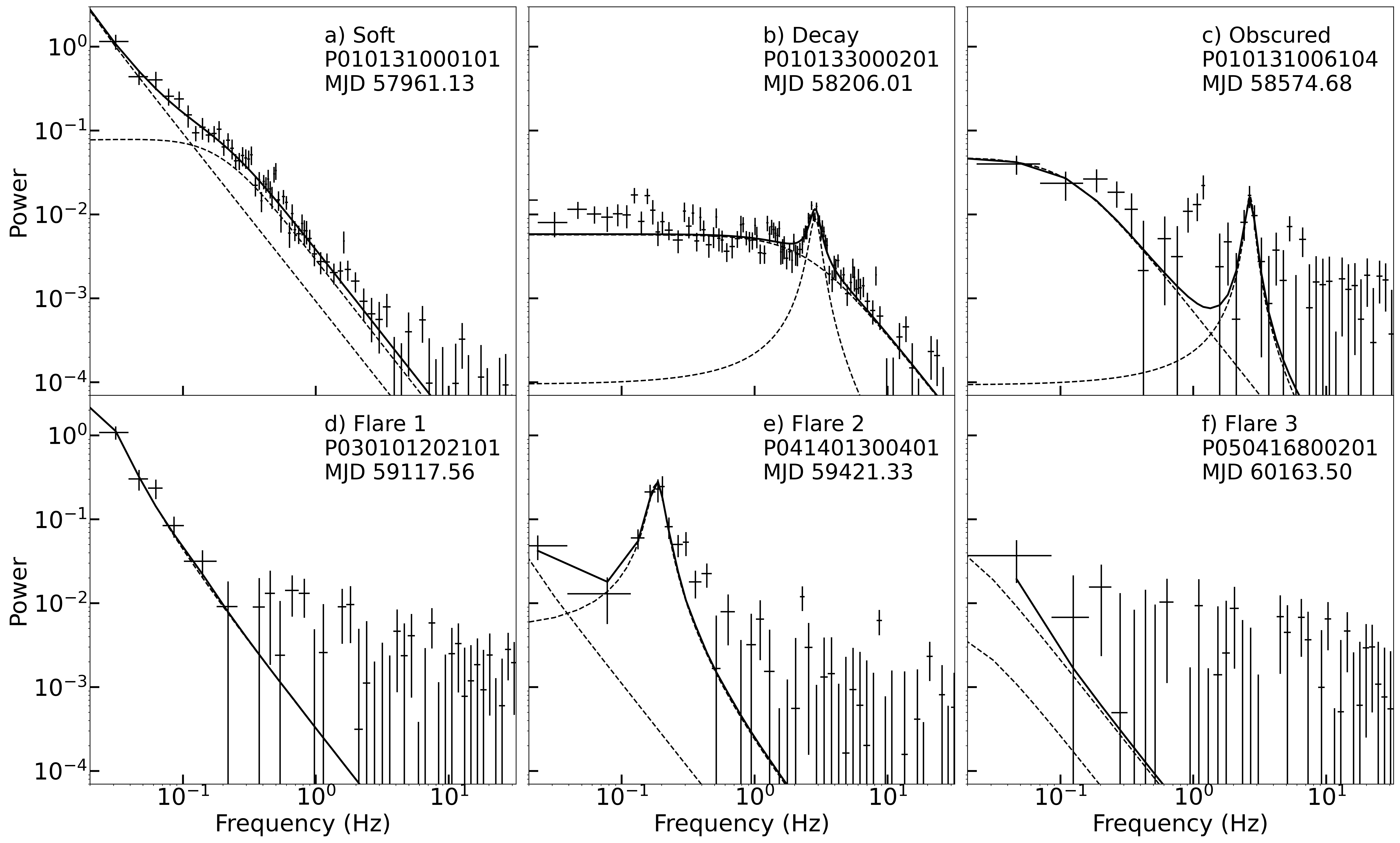}
    \caption{Representative LE power spectra for the three main evolutionary phases (a-c) and the three flares (d-f). QPOs with frequencies above 1~Hz are commonly detected during the decay phase (panel b) and occasionally during the obscured state (panel c), whereas QPOs with frequencies around 0.2~Hz appear in the second flare (panel e).}
    \label{fig:6PDS}
\end{figure*}

QPOs with frequencies above 1 Hz are commonly observed in GRS~1915+105. We present the properties of these $>1$ Hz QPOs in Figure~\ref{fig:evolve_QPO}, including their centroid frequencies, full width at half maximum (FWHM), and fractional rms. To quantitatively demonstrate the reliability of the QPO signals, we also calculated their significance, shown in the bottom panel of Figure~\ref{fig:evolve_QPO}. Based on these significance results, all points with values below $3\sigma$ are plotted in grey for all the panels. Nevertheless, these low-significance signals could still be real. For instance, the QPOs around MJD 58600 have a significance of only $\sim$ 2 $\sigma$ in our analysis, yet they have also been reported in NICER observations, with even harmonics detected \citep{Koljonen2021A&A...647A.173K, Zhou2025A&A...694A.104Z}. The lower significance in Insight-HXMT data is likely due to its lower signal-to-noise ratio and higher background level compared to NICER. The QPO frequencies show no significant energy dependence. However, the QPO frequencies in LE, ME, and HE rise from approximately 2~Hz to 6~Hz and then gradually decreases back to around 2~Hz, indicating that a full state transition occurred during the decay phase, which has not been previously reported. An example of the LE PDS from this stage is shown in panel b of Figure~\ref{fig:6PDS}. For comparison, one representative LE power spectrum is presented in Figure~\ref{fig:6PDS} for each of the three main evolutionary phases (panels a–c) and the three flares (panels d–f). QPOs with frequencies above 1~Hz are commonly detected during the decay phase (panel b) and occasionally during the obscured state (panel c), whereas QPOs around 0.2Hz are observed during the second flare (panel e). No significant QPOs are detected in the PDS of the other phases. Meanwhile, the QPO widths remain nearly stable with a slightly decrease. The decrease is more apparently in HE and ME than that in LE, which is also evident in the fractional rms. The fractional rms of LE remains basically constant. During the decay phase, the fractional rms of HE and ME is higher than LE, suggesting a corona origin.

\begin{figure}
    \begin{minipage}{0.48\textwidth}
		\includegraphics[width=1\textwidth]{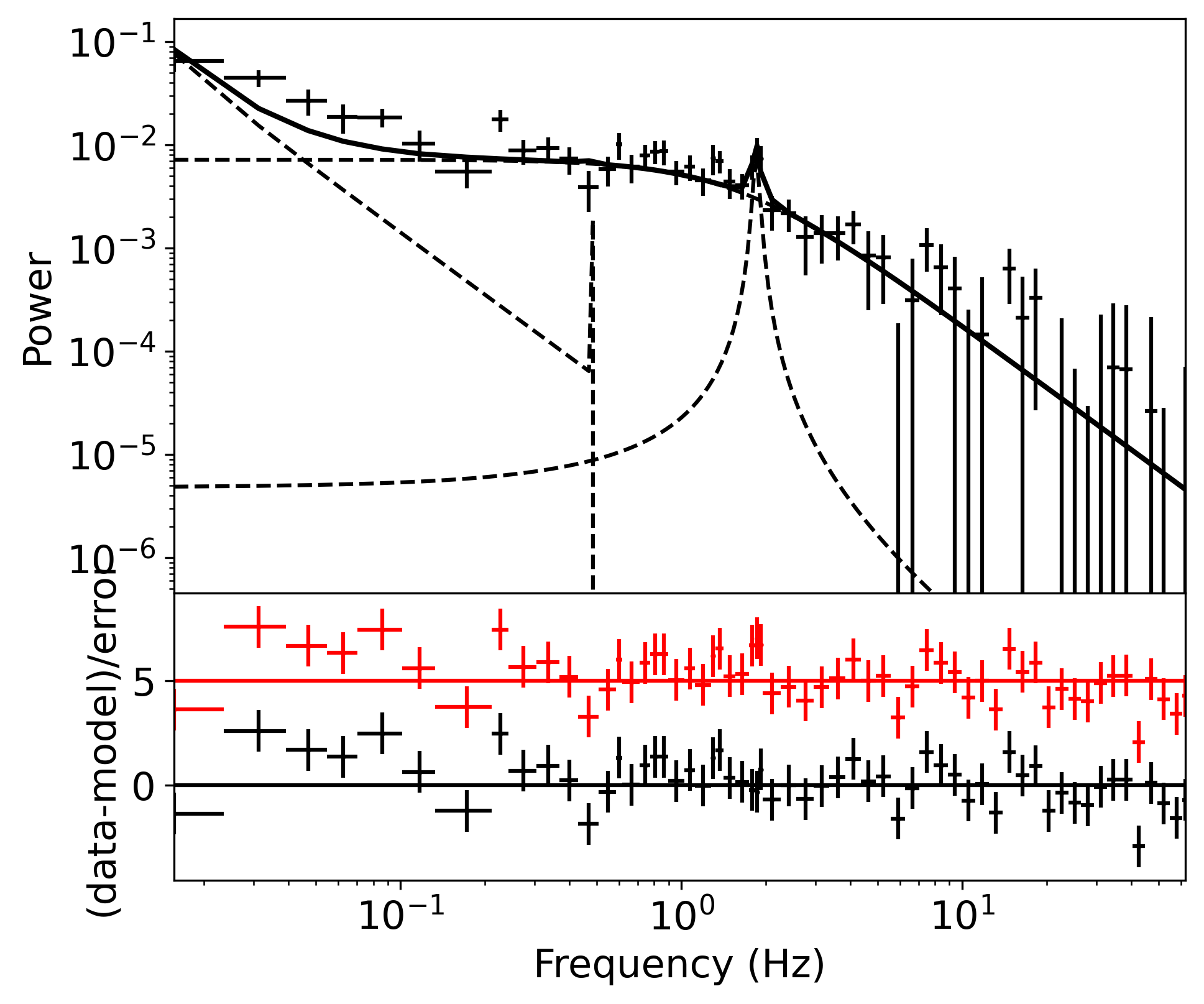}
	\end{minipage}
	\begin{minipage}{0.48\textwidth}
		\includegraphics[width=1\textwidth]{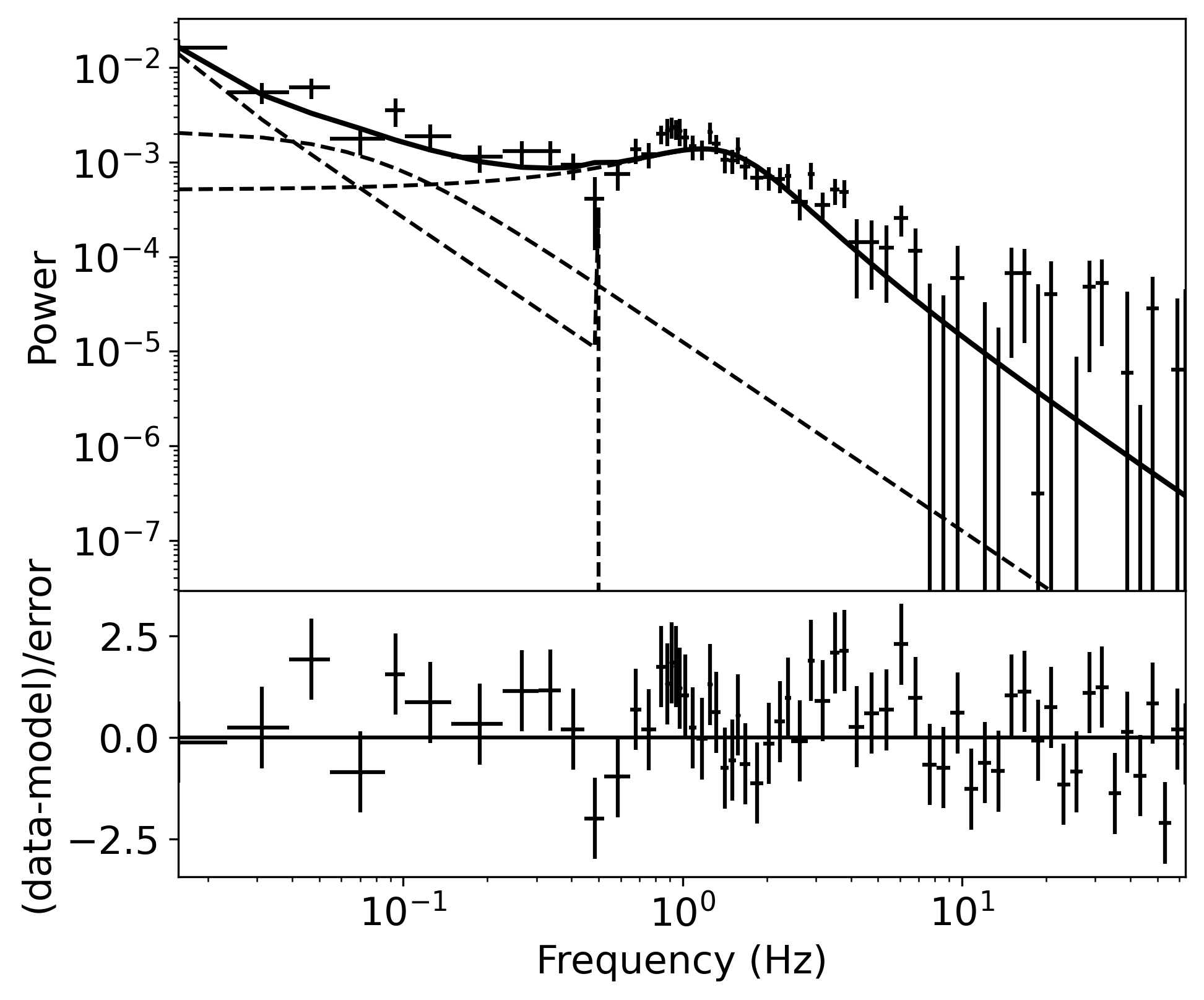}
	\end{minipage}
    \caption{{\bf Top panel}: The ME PDS for observation P010131000202 (MJD 58055). The residuals in black show the best fitted results, while residuals in red indicate the fitting results after subtracting the QPO component (shifted upward by 5 for clarity). A distinct peak is visible around 2~Hz.  {\bf Bottom panel}: The LE PDS for the same observation. The broad peak around 1~Hz could not be well fitted with a Lorentzian.}
    \label{fig:type-B QPO}
\end{figure}

We note that around MJD 58055 in Figure~\ref{fig:evolve_QPO}, i.e. during the soft state and just prior to a small sudden flux drop, two possible QPOs are detected exclusively in the ME band. In observation P010131000202 (MJD 58055), the ME PDS reveals a QPO at $\sim$ 1.82 $\pm$ 0.03 Hz with a quality factor $Q = 21$ and an rms amplitude of $\sim$ 4\%. The corresponding PDS is shown in the top panel of Figure~\ref{fig:type-B QPO}. A band limited red noise is evident at lower frequencies, suggesting that this QPO signal may be a type-B QPO. Although the centroid frequencies are lower than the typical range, \cite{Motta2011MNRAS.418.2292M} have shown that type-B QPOs can occur between 1--7~Hz. However, no corresponding QPO is detected in the LE or HE bands. The LE PDS is shown in the bottom panel of Figure~\ref{fig:type-B QPO}, with only a very broad feature at the same frequency. Due to high background contamination, the HE PDS is not shown. A similar weak QPO feature is also observed in observation P010131000305 (MJD 58056), with a centroid frequency of 2.75 $\pm$ 0.06 Hz and a Q factor of 13. These two observations exhibit similar timing properties, with LE count rate at $\sim 584 cts/s$, hardness ratio (3--10 keV / 1--3 keV) of 3.73, and a second hardness ratio (10--35 keV / 1--3 keV) of 1.18. As shown in the top panel of Figure~\ref{fig:hr}, they are located within the upper branch of the soft state. However, not all data points in this region show detectable QPOs, possibly due to their transient nature or intrinsically low amplitudes. However, we note that the significance of these two QPOs is only $\sim$ 2 $\sigma$. Although a significance below 3 $\sigma$ does not necessarily mean that the QPOs are not real, they may also represent artifacts from the ME detector and should be interpreted with caution. Unfortunately, \textit{NICER} observations are only available before 2017-10-17 and after 2017-10-31, leaving this interval unobserved.

\subsubsection{QPOs below 1~Hz}

In addition to the $>1$~Hz features, GRS~1915+105 also exhibits prominent variability in the range of sub-Hertz or millihertz (mHz). \cite{Kong2024A&A...686A.211K} reported QPO features around 0.2~Hz during the second flare using observations from the Neutron Star Interior Composition Explorer (NICER). These features are also very prominent in the Insight-HXMT data, as shown in panel e of Figure~\ref{fig:6PDS} for the rms-normalized PDS. The $\sim$ 0.2~Hz QPO is strong in the LE band, weaker in the ME band, and barely visible in the HE band. The disappearance in the HE band may be attributed to either the intrinsic energy dependence of the QPO signal or the high background level in it.

Besides the $\sim$ 0.2~Hz QPO mentioned above, a $\sim$ 0.02~Hz QPO has also been detected after the decay phase by \textit{AstroSat}, \textit{NuSTAR} and \textit{NICER} \citep{Athulya2023MNRAS.525..489A}. However, identifying such low frequency signals in the PDS is challenging due to limited frequency resolution and the presence of strong noise fluctuations in this range. Additionally, sub-Hertz and mHz QPOs in GRS~1915+105 are sometimes weak and transient, appearing only for short durations within individual observations or GTIs. These signals may not persist throughout the entire exposure, making them difficult to detect and characterize using traditional PDS fitting techniques. Thus to overcome these limitations, we employed wavelet analysis instead of regenerating PDS with longer segments (e.g. 128 s or longer), which require sufficiently long exposure for reliable results. Wavelet analysis provides better sensitivity to non-stationary and time-localized signals, allowing us to identify and track the evolution of these low-frequency features more effectively. In particular, wavelet analysis preserves both time and frequency resolution, making it well suited for capturing transient and evolving QPO behavior.

\begin{figure}
    \centering
    \begin{minipage}{0.48\textwidth}
		 \includegraphics[width=\columnwidth]{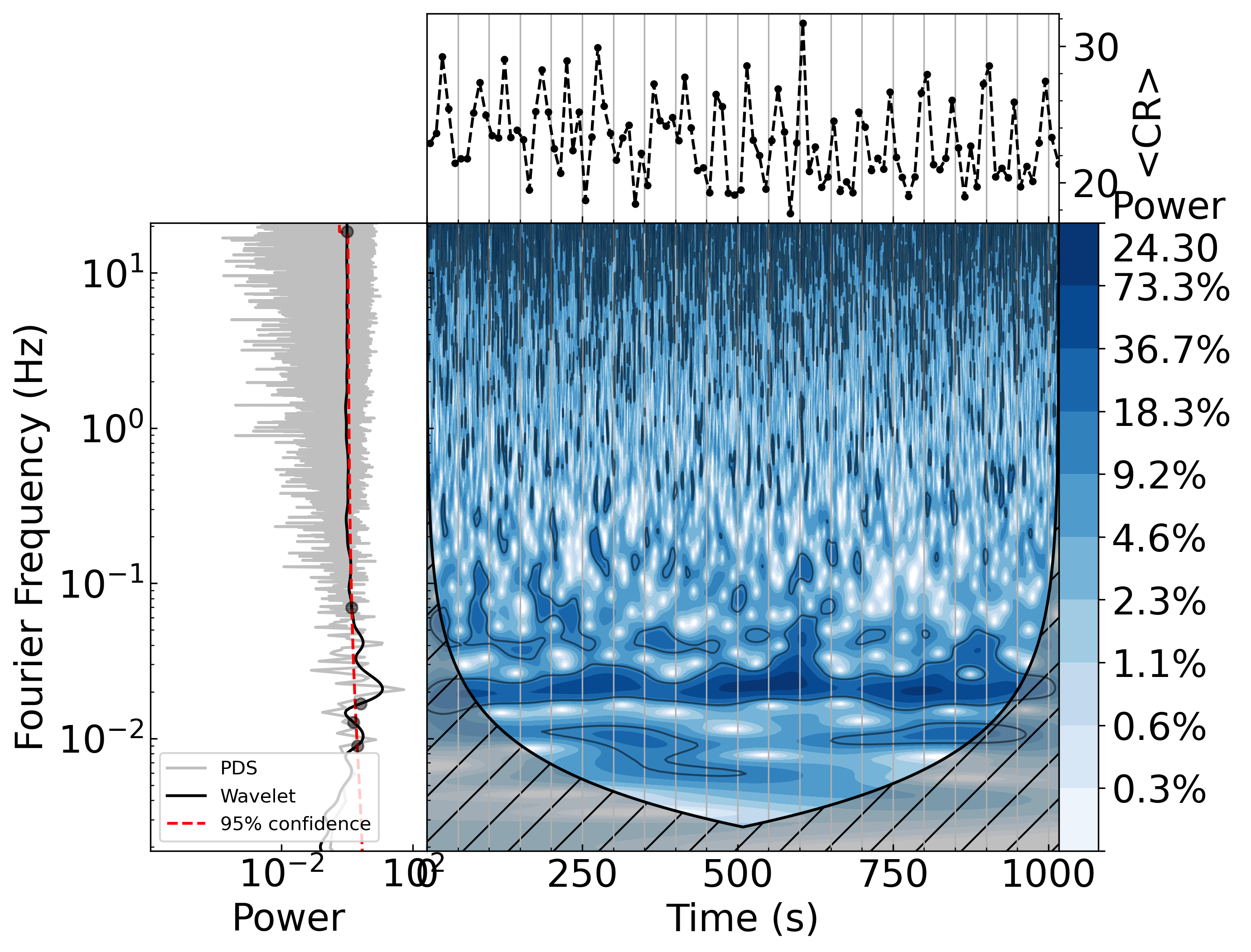}
	\end{minipage}
   \begin{minipage}{0.48\textwidth}
		 \includegraphics[width=\columnwidth]{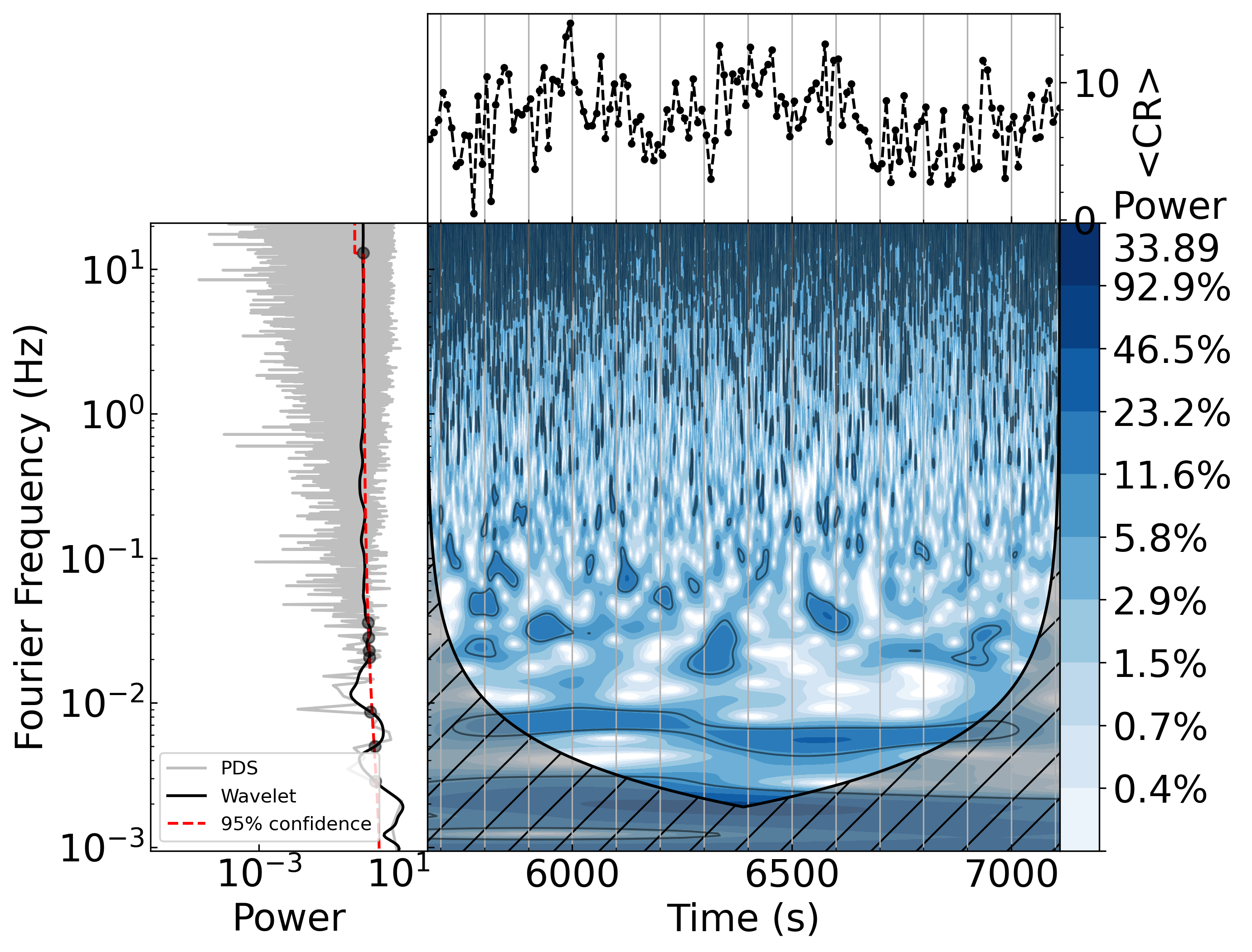}
	\end{minipage}
    \caption{{\bf Top panel: } Wavelet analysis of observation P030101202401 (MJD~59124) in the LE band. The left sub-panel shows the global wavelet power spectrum (black curve), with the corresponding PDS overlaid in grey for comparison. The red dashed line indicates the 95\% confidence threshold, and intersections between this threshold and the global spectrum are marked with filled dark circles. The right sub-panel shows the wavelet contour map (local wavelet power spectrum), where regions exceeding the 95\% confidence level are outlined in black. The cone of influence, indicating areas affected by edge effects, is shown as a grey hatched region at the bottom. The color bar of the contour map indicating power levels is shown on the right side. The top sub-panel shows the averaged count rate of every 10~seconds. A $\sim$0.02~Hz QPO can be identified in both the global and local wavelet power spectra. {\bf Bottom panel: } Wavelet analysis of observation P030101201702 (MJD~59110) in the ME band. A $\lesssim$0.01~Hz QPO can be identified in both the global and local wavelet power spectra.}
    \label{fig:wavelet_0.01Hz}
\end{figure}

Figure~\ref{fig:wavelet_0.01Hz} shows an example of a $\sim$ 0.01~Hz QPO during Flare 1. Similar results have been reported by \cite{Athulya2023MNRAS.525..489A} and \cite{Kong2024A&A...686A.211K} with NICER data during this flare. We find that this mHz QPO appear is strong in observation P030101200101 (MJD 59091), but becomes weak or intermittent after P030101201101 (MJD 59102), and reapparing in P030101201301 (MJD 59105). At that time, however, the mHz QPO is heavily contaminated by other low-frequency noise $\leq$ 0.1~Hz, likely due to significant variability in the light curve. The QPO signal becomes more distinct in P030101202301 (MJD 59121), and disappears after P030101202601 (MJD 59128). Throughout this interval, the mHz QPO is generally not apparent in the ME band, though it appears intermittently between MJD 59105 and MJD 59121, which is the time when LE data becomes noisy around 0.01~Hz. An example is shown in the bottom panel of Figure~\ref{fig:wavelet_0.01Hz}. Due to the high background level in the HE data during this period, no significant QPO signals are observed in it.

\begin{figure}
    \centering
    \includegraphics[width=\columnwidth]{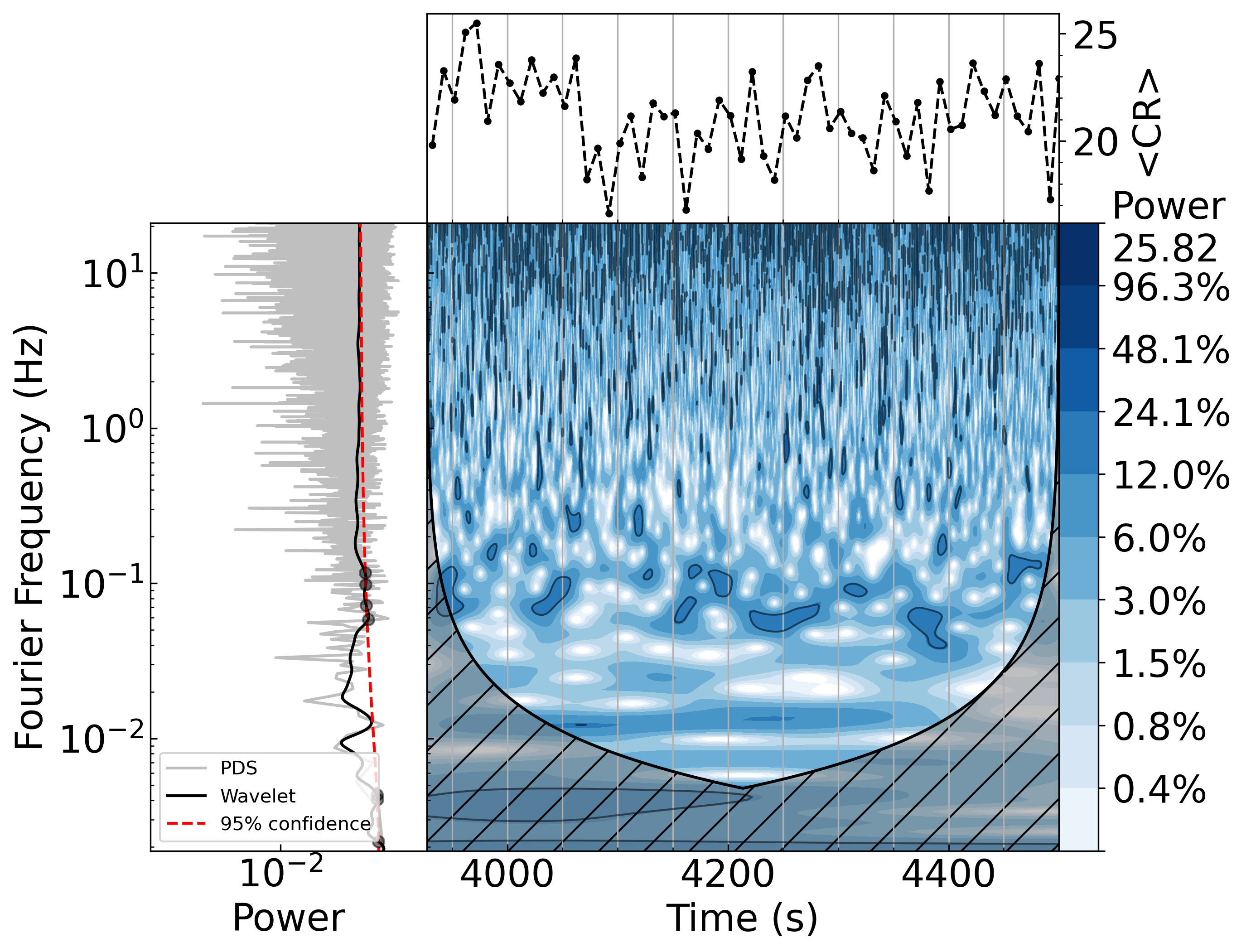}
    \caption{Wavelet result for the weak mHz QPO in the LE data of observation P050416800701 (MJD~60168) during Flare 3. A QPO around 0.1 Hz appears intermittently and exhibits noticeable frequency drift. The labeling and symbols are consistent with those in Figure~\ref{fig:wavelet_0.01Hz}.}
    \label{fig:wavelet_flare3_0.1Hz}
\end{figure}

During Flare 3 and the preceding mini-burst(s), a QPO feature $\sim$ 0.2~Hz is weak but still visible, with its frequency ranging from around 0.1 to 0.4~Hz. Figure~\ref{fig:wavelet_flare3_0.1Hz} presents an example of this weak signal in the LE data of observation P050416800701 (MJD~60168). In this figure, a QPO feature at $\sim$0.1~Hz is observed intermittently, accompanied by noticeable frequency drift, suggesting a non-stationary origin or changes in the accretion flow geometry. This QPO signal is not detected in the ME and HE data.

\subsection{Spectral analysis}

Spectral analysis provides complementary insights into the physical conditions of the accretion flow and the nature of the emitting regions. By studying the evolution of spectral parameters across different states and phases, we could better understand the underlying mechanisms of the timing features and energy-dependent behavior.

\subsubsection{Spectral modelling}

GRS 1915+105 is known to exhibit ionized disk winds \citep{Lee2002ApJ...567.1102L,Martocchia2006A&A...448..677M,Koljonen2021A&A...647A.173K,Zhou2025A&A...694A.104Z}. Following \cite{Zhou2025A&A...694A.104Z}, we adopt the \textit{zxipcf} model to describe the ionized absorption component. We fix the covering fraction at 1 and the redshift at 0. We choose \textit{xillverCp} and \textit{relxillCp} to characterize the distant reflector and the reflection from the accretion disk, respectively \citep{Garcia2013ApJ...768..146G,Garcia2014ApJ...782...76G,Dauser2014MNRAS.444L.100D}, with their soft excess corrected by \textit{nthratio}. Although the \textit{diskbb} component is sometimes uncessary or only marginally improves the fit in previous HXMT and NICER studies during the soft state and decay phase \citep[e.g.,][]{Liu2022ApJ...933..122L,Zhou2025A&A...694A.104Z}, we retain it here for a more comprehensive and consistent modeling. Thus, we start with the following spectral model:

\begin{multline}
    \texttt{Constant} \times \texttt{TBabs} \times \texttt{zxipcf} \times (\texttt{nthComp} + \texttt{diskbb} \\ + \texttt{nthratio} \times (\texttt{relxillCp} + \texttt{xillverCp})).
\end{multline}

We refer to it as Model A hereafter. To ensure physical self-consistency and reduce parameter degeneracy, we linked several parameters together. The photon index $\Gamma$ and electron temperature $kT_{\mathrm{e}}$ in both \texttt{relxillCp} and \texttt{nthratio} were tied to those in \texttt{nthComp}, assuming that they all originate from the same corona. The seed photon temperature $kT_{\mathrm{bb}}$ in both \texttt{nthComp} and \texttt{nthratio} were linked to the inner disk temperature $T_{\mathrm{in}}$ in the \texttt{diskbb} component. The two reflection components (\texttt{relxillCp} and \texttt{xillverCp})) share the same parameters except for their normalization. In addition, several parameters were fixed during the fitting. The inclination angle was fixed at $i = 60^\circ$ \citep{Reid2014ApJ...796....2R}, and the black hole spin was fixed at 0.98 \citep{Miller2013ApJ...775L..45M,Reid2014ApJ...796....2R}. The reflection fraction in both \texttt{relxillCp} and \texttt{xillverCp} was fixed at $-1$ as required when using \texttt{nthratio}. Given the difficulty of constraining the hydrogen column density $N_{\mathrm{H}}$ of the interstellar medium during the obscured state, it was fixed at $6 \times 10^{22} cm^{-2}$, consistent with previous studies \citep{Zhou2025A&A...694A.104Z}. While \cite{Neilsen2020ApJ...902..152N} reported an increase in absorption during the obscured state and suggested that an additional neutral absorber may be required, \cite{Kong2024A&A...686A.211K} found that an extra \texttt{pcfabs} component was not necessary during the second flare. Since our primary goal is to study the flares during the obscured state, we maintain the use of Model A for consistency, and the fits obtained are acceptable. The iron abundance $A_{\mathrm{Fe}}$ was fixed at the solar value.

\begin{figure}
    \centering
    \includegraphics[width=\columnwidth]{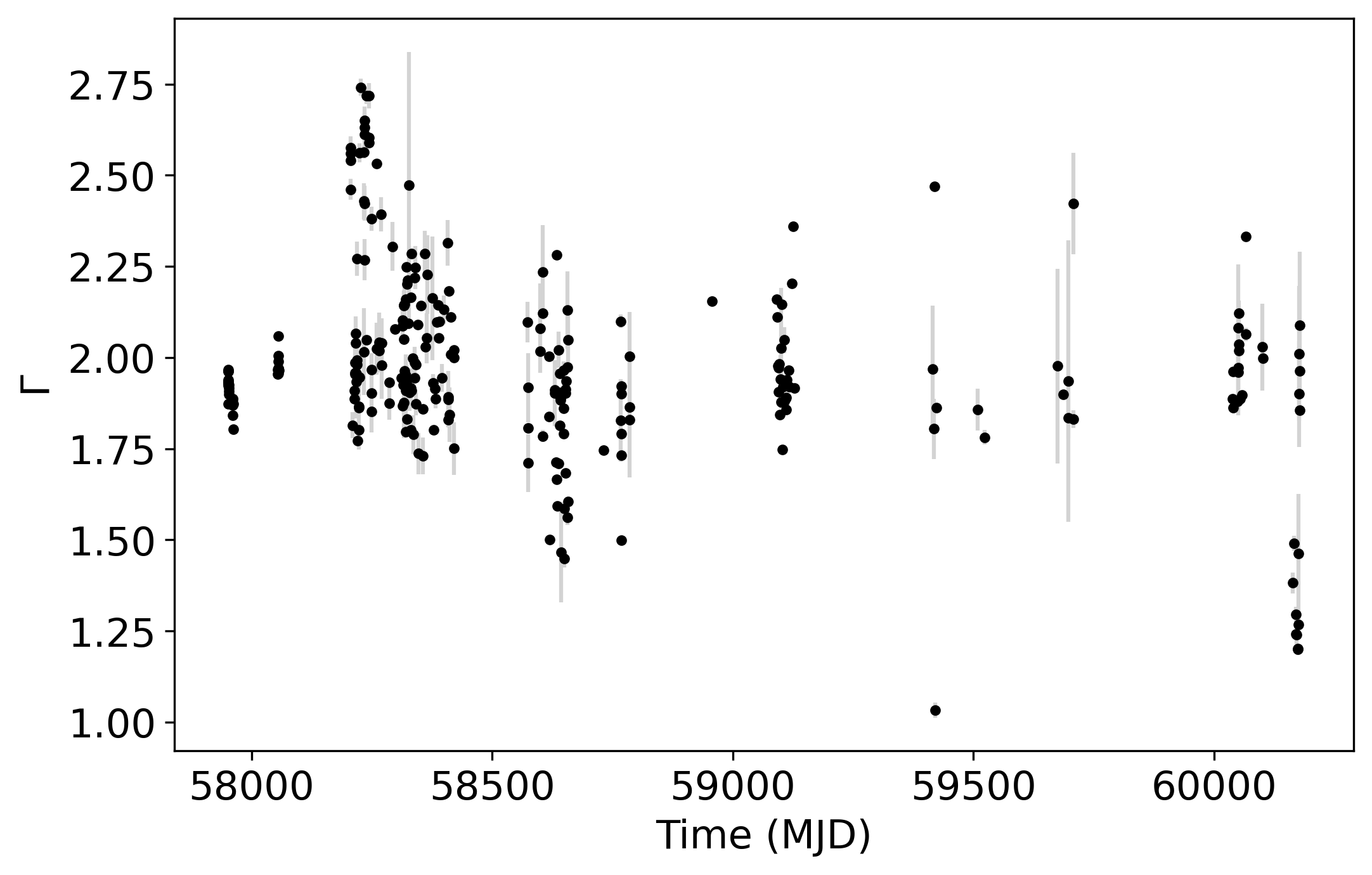}
    \caption{Evolution of the photon index $\Gamma$ derived from spectral fitting with Model A. Data points are shown with light grey error bars.}
    \label{fig:evolve_relxillGamma}
\end{figure}

Inclusion of HE spectra in Model A fitting was unsuccessful due to the insufficient number of energy bins (only around 13 bins). Therefore, we use only the LE and ME spectra for this model. The fitting are generally acceptable, with no significant structures observed in the residuals. Most of the chi-square values are within the range of $\sim$ 800 -- 1100 with a degrees of freedom (dof) of 984. However, we notice that the photon index $\Gamma$ becomes highly scattered in the decay phase, making this fitting unreliable. In Figure~\ref{fig:evolve_relxillGamma}, we present the evolution of $\Gamma$ as derived from Model A. \cite{Koljonen2021A&A...647A.173K} and \cite{Zhou2025A&A...694A.104Z} both showed a clear decreasing trend in the decay phase, but the trend is barely seen in our results. This discrepancy may be attributed to the limited capability of Insight-HXMT data to simultaneously constrain the large number of free parameters in the reflection model, especially in faint states.

Thus, we attempted to simplify the spectral model. As demonstrated in \cite{Liu2022ApJ...933..122L}, the soft state spectra can be adequately fitted using a simple model: $\texttt{TBabs} \times \texttt{pcfabs} \times (\texttt{nthComp} + \texttt{Gaussians})$. Following this approach, we simplified our model accordingly as

\begin{multline}
    \texttt{Constant} \times \texttt{TBabs} \times \texttt{zxipcf} \times (\texttt{nthComp} + \texttt{diskbb}),
\end{multline}
and Gaussian components are added around 6.6 -- 6.8~keV when necessary. We refer to this as Model B hereafter. Meanwhile, all three energy bands could be fitted simultaneously using this model.

\begin{figure*}
    \centering
    \includegraphics[width=\textwidth]{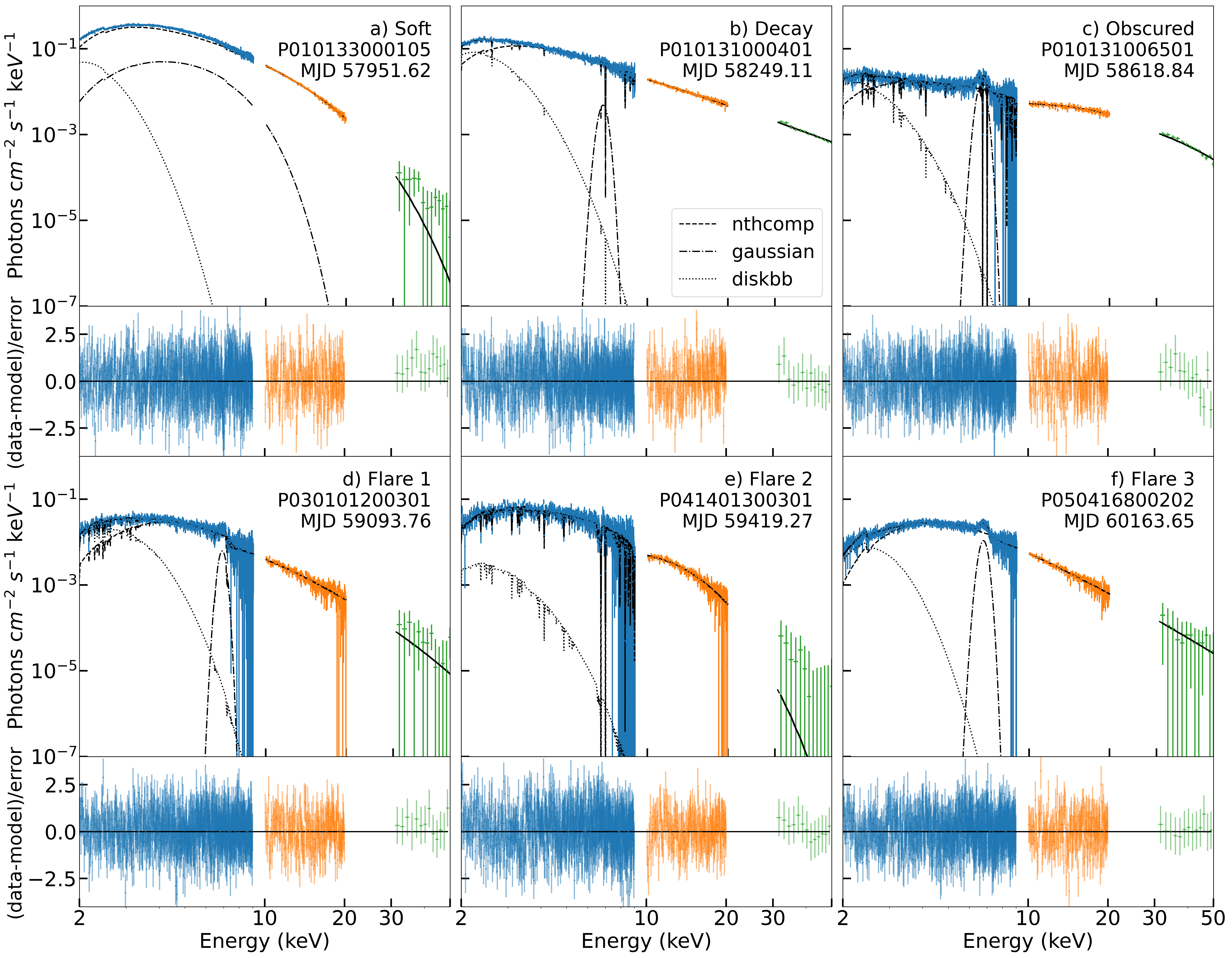}
    \caption{One representative energy spectrum is shown for each of the three main evolutionary phases (a-c) and the three flares (d-f) using Model B. The blue, orange and green lines indicate the LE, ME and HE data, respectively. The nthcomp, gaussian, and diskbb components are shown with dashed, dash-dotted, and dotted lines, respectively, as indicated in panel b.}
    \label{fig:specFitting_Gau}
\end{figure*}

This simplified model generally produced acceptable fits with most reduced chi-square values falling in the range of $\sim$ 850 -- 1050 for 998 dof, except during the soft state at the beginning of the dataset. Figure~\ref{fig:specFitting_Gau} presents one representative energy spectrum for each of the three main evolutionary phases (panels a-c) and the three flares (panels d-f). For those early soft state observations, a very broad Gaussian component had to be added to achieve an acceptable fit. As shown in panel a of Figure~\ref{fig:specFitting_Gau}, this Gaussian component appears at a peak energy of around 3~keV with a width of $\sim$ 3~keV. Such unusual broad emission near 3~keV suggests that an additional reflection component may be required at this stage, although other possible origins (e.g. a secondary Comptonisation continuum) cannot be ruled out. Removing this Gaussian component increases the $chi^2$/dof from $\sim$1000-1100/998 to $\sim$1100-2000/1001. Therefore, we retained this component in the final model. For reliability, we also compared these soft-band results with the LE + ME fittings obtained using Model A, and found that the parameters are generally consistent.

\subsubsection{Spectral results}

\begin{figure*}
    \centering
    \includegraphics[width=\textwidth]{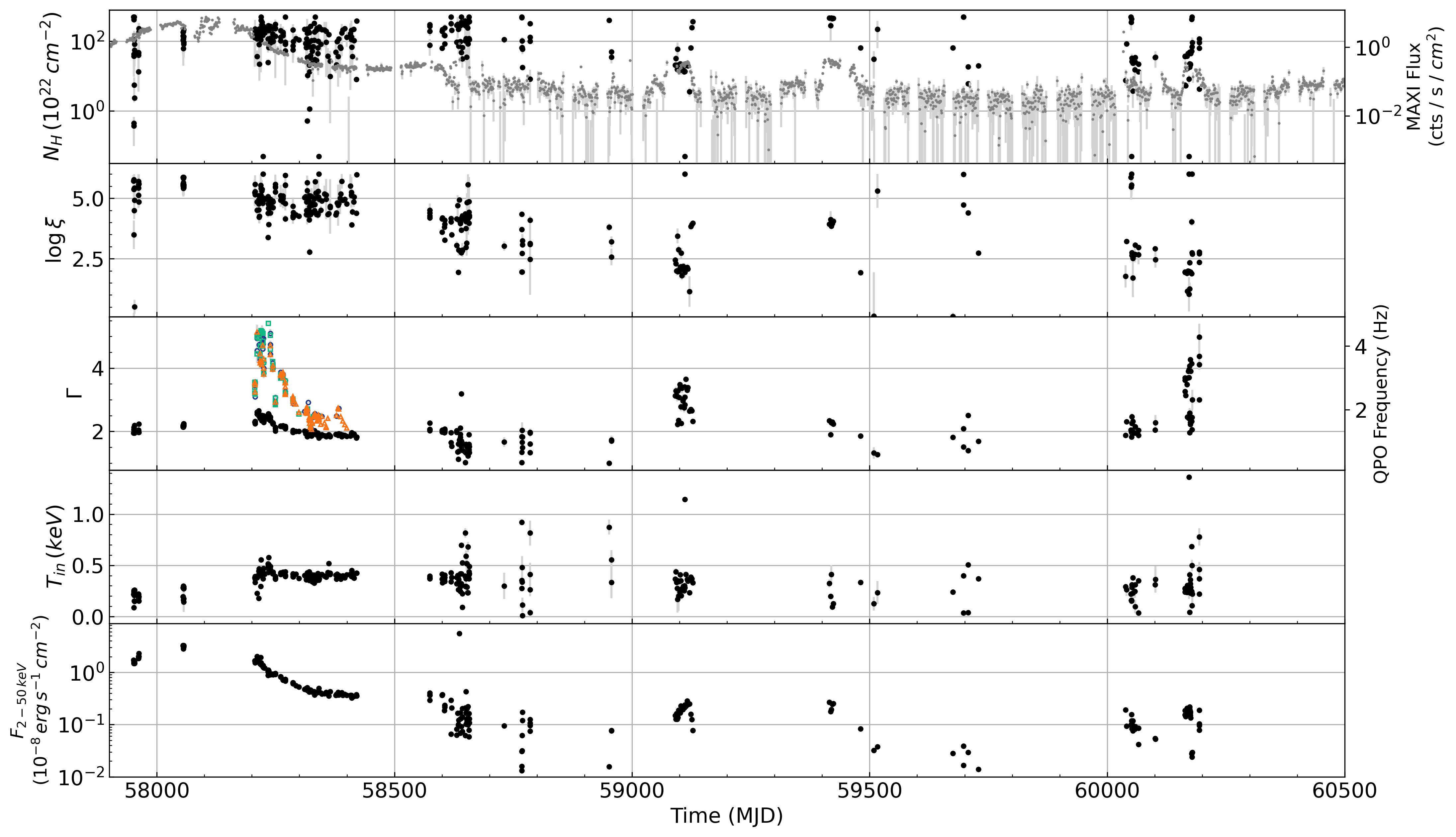}
    \caption{Spectral fitting results from 2--50 keV using Model B ($\texttt{Constant} \times \texttt{TBabs} \times \texttt{zxipcf} \times (\texttt{nthComp} + \texttt{diskbb})$). From top to bottom: column density of the ionized absorber $N_{\mathrm{H}}$ (black points) with the MAXI flux (grey points); ionization parameter $\log \xi$; photon index $\Gamma$ of the power-law component, with QPO frequencies above 3$\sigma$ significance plotted as blue circles (LE), green squares (ME), and orange triangles (HE); inner disk temperature $T_{\mathrm{in}}$; and the absorbed 2--50 keV flux density. All data points are shown with light grey error bars.}
    \label{fig:evolve_specParams}
\end{figure*}

The final spectral results in Figure~\ref{fig:evolve_specParams} show the evolution of key parameters, including the column density of the ionized absorber $N_{\mathrm{H}}$, ionization parameter $\log \xi$, photon index $\Gamma$, inner disk temperature $T_{\mathrm{in}}$ and the absorbed 2--50 keV flux density $F_{\mathrm{2-50 keV}}$. The electron temperature $kT_{\mathrm{e}}$ is not shown, as it cannot be well constrained, especially in the decay phase.

During the initial soft state, the photon index $\Gamma$ remains stable at approximately 2. As the source enters the decay phase, $\Gamma$ first increases to $\sim$ 2.6 and then gradually decreases to $\sim$ 1.8. This decrease is broadly consistent with the flux decay (i.e., an exponential decay followed by a linear decay) reported by \cite{Koljonen2021A&A...647A.173K}. Specifically, \cite{Koljonen2021A&A...647A.173K} found that $\Gamma$ decreased from $\sim$2.7 to $\lesssim2.0$, which agrees well with our trend. In comparison, \cite{Zhou2025A&A...694A.104Z} reported $\Gamma \sim 3.4$ in the soft state, decreasing to above 2 in the decay phase, slightly higher than both our values and those in \cite{Koljonen2021A&A...647A.173K}. \cite{Liu2022ApJ...933..122L} measured $\Gamma$ in the range 2.3 -- 2.6 for the soft state using Insight-HXMT data, values that are slightly higher than ours but notably lower than those of \cite{Zhou2025A&A...694A.104Z}.

For $T_\mathrm{in}$, we obtain values around 0.3~keV in the soft state. At the beginning of the decay phase, $T_\mathrm{in}$ rises rapidly from $\sim$ 0.2 to $\sim$ 0.4~keV and then remains nearly constant. These results are broadly consistent with \cite{Zhou2025A&A...694A.104Z} during the decay phase (below 0.5~keV), but significantly lower in the soft state, where they reported $\sim$1~keV. Even higher values were reported by \cite{Neilsen2018ApJ...860L..19N}. We tested Model A and obtained similar low $T_\mathrm{in}$ values. This discrepancy may arise from our fixed interstellar column density ($N_{\mathrm{H}} = 6 \times 10^{22}$ cm$^{-2}$), which is lower than the fitted $N_{\mathrm{H}}$ in \cite{Liu2022ApJ...933..122L}. Meanwhile, \cite{Liu2022ApJ...933..122L} noted that adding a \texttt{diskbb} component improved the fit only marginally ($\Delta\chi^2 \sim 3$), suggesting that the disk emission is intrinsically weak. In any case, $T_\mathrm{in}$ should remain low during the soft state.

The ionization parameter $\log \xi$ generally stays high ($> 5$), with occasional lower points accompanied by large uncertainties during the initial soft state. \cite{Zhou2025A&A...694A.104Z} reported no strong ionized absorption features during the decay phase based on NICER data. However, although the ionized absorption decreases slightly in the decay phase, it remains strong before showing a notable decline as the source enters the obscured state. Meanwhile, the absorption column density in the soft state and the decay phase cannot be well constrained because the winds are highly ionized. When $\xi$ is large, the absorbing material becomes nearly transparent to X-rays, weakening absorption features and making $N_{\rm H}$ of the ionized absorber difficult to determine reliably. Nevertheless, the column density is mainly concentrated at $\sim 10^{24} cm^{-2}$.

During the obscured state, the low flux makes spectra fitting challenging, resulting in relatively scattered parameter values. During the first flare, $\log \xi$ decreased to $\lesssim$ 2, while $\Gamma$ increased to $\sim$ 3.5, and $T_{\mathrm{in}}$ also showed a slight decline. Although we lack sufficient observations for the second flare, it can be seen that at its peak flux, $\Gamma$ and $\log \xi$ show an opposite trend compared to Flare 1, with $\log \xi$ increased while $\Gamma$ decreased. Meanwhile $T_{\mathrm{in}}$ remained low. In the third flare, the flux in LE band is generally lower than that of the first two flares in both Insight-HXMT and MAXI, but the ME and HE bands show relatively higher flux. The results are more scattered possibly because this flare underwent a more complex outburst, as indicated by the hardness ratio evolution in Figure~\ref{fig:evolve_lc} and Figure~\ref{fig:hr}. Nonetheless, the parameter evolution of Flare 3 resembles that of Flare 1, with $\log \xi$ decreasing and $\Gamma$ increasing, suggesting a generally weaker ionized absorption. Prior to Flare 3, a mini-burst or a series of bursts appears, exhibiting a trend more similar to Flare 2, characterized by a lower $\Gamma$ and a higher $\log \xi$. Throughout the entire obscured state, $T_{\mathrm{in}}$ generally remains low.

\section{Discussion}
\label{sec:discussion}

Previous studies on GRS 1915+105 have extensively explored its variability classes, spectral state transitions, QPOs, ionized disk winds, and re-brightening since the source became faint in 2018 \citep{Shi2023MNRAS.525.1431S,Athulya2023MNRAS.525..489A,Liu2021ApJ...909...63L,Liu2022ApJ...933..122L}. Spectral-timing research was also conducted to give a view of its variability \citep[e.g.][]{Koljonen2021A&A...647A.173K,Zhou2025A&A...694A.104Z}. However, none of these works have reported the increasing trend of QPO frequency shortly after MJD 58200, nor have they addressed the properties of the third flare in detail. The Insight-HXMT observations, with their broad energy coverage, help bridge this observational gap by providing crucial coverage of the source during its fading phase.

\subsection{The QPO evolution in the decay phase}

Several previous studies have examined the QPO frequency evolution in GRS 1915+105 during its decay phase. For instance, \cite{Koljonen2021A&A...647A.173K} utilized NICER data and found that the QPO frequency decreased steeply during the exponential decay phase, followed by a smoother decrease in the subsequent linear decay, with the transition occurring around MJD 58330. This evolution was also reported by \cite{Zhou2025A&A...694A.104Z}, who primarily used NICER data as well. However, in both studies, the QPO frequency measurements began at $\sim$ MJD~58238, leaving the earlier phase unexamined. Although \cite{Liu2021ApJ...909...63L} analyzed Insight-HXMT observations, their research data start from MJD 58216, thus also omitting this earlier phase. Meanwhile, our Insight-HXMT analysis begins earlier at MJD~58206 (observation P0101330002), thus covering the previously unexplored phase when the QPO frequency was still increasing.

\begin{figure}
    \centering
    \includegraphics[width=\columnwidth]{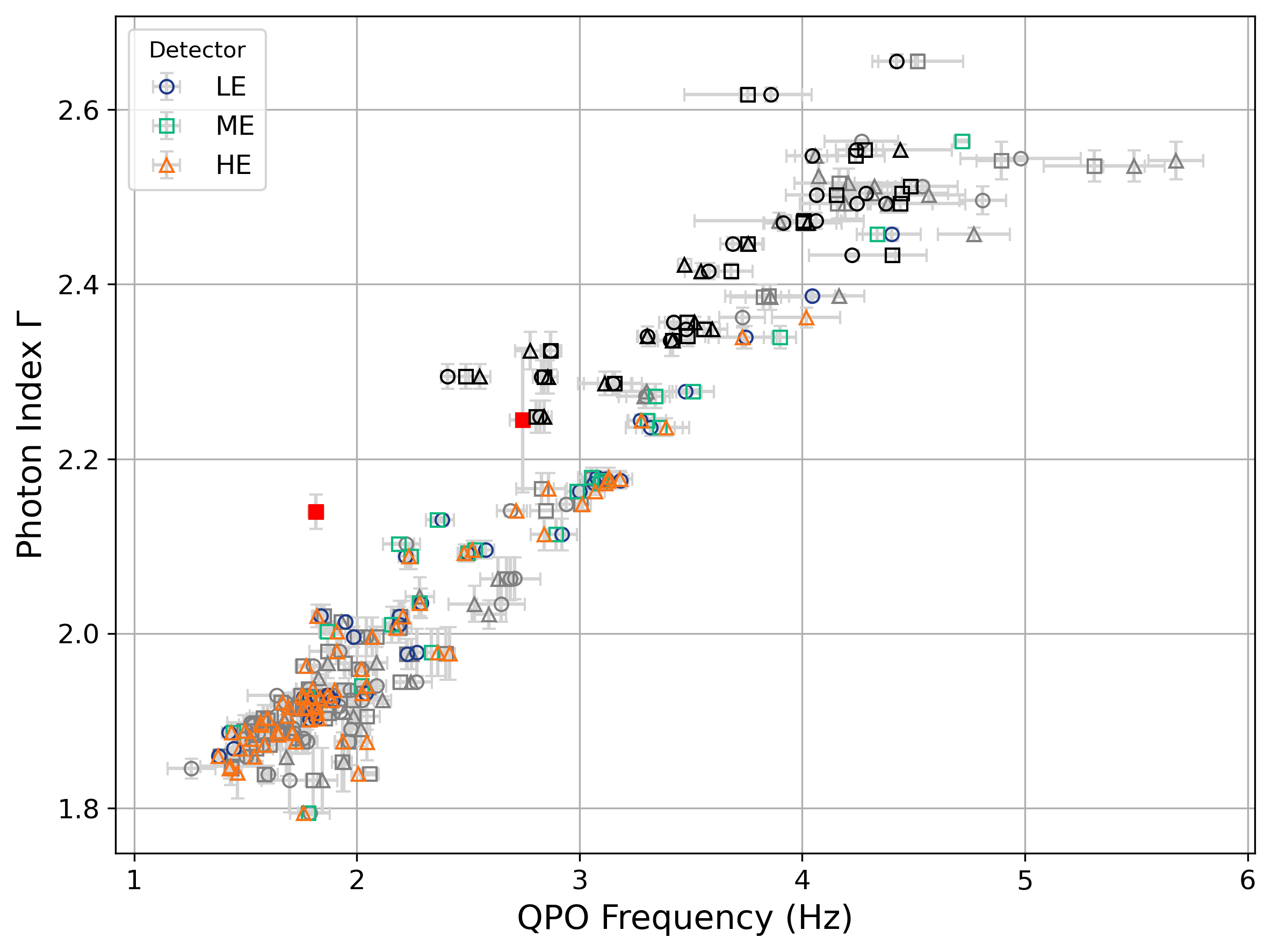}
    \caption{Photon index versus QPO centroid frequency. Data points with significance $\geq$3 are plotted in blue circles (LE), green squares (ME), and orange triangles (HE), while those with significance $<$3 are shown in grey. Black symbols represent observations within MJD~58200–-58230, corresponding to the rising phase of the QPO frequency. The two filled red squares show the suspected type-B QPOs. All data points are shown with light grey error bars.}
    \label{fig:QPO_Gamma_relation}
\end{figure}

Such behavior suggests that the source completed a full spectral state transition cycle. The QPO frequency rising branch is typically associated with the transition from the hard state to the intermediate state. During this phase, the inner disk moves inward, increasing the disk temperature and soft photon flux, while the corona becomes less dominant, which leads to spectral softening. As the QPO frequency reaches its maximum and begins to fall, the source likely transits back toward the hard state with decreasing $\Gamma$. Such an evolution of QPO frequency, together with its correlated spectral behavior, is commonly seen in black hole X-ray binaries. \cite{Zhou2025A&A...694A.104Z} reported a positive correlation between QPO frequency and $\Gamma$ for observations after MJD 58230. When the rising branch between MJD 58206 and 58230 is included (Figure~\ref{fig:QPO_Gamma_relation}), the correlation is still evident.

\begin{figure}
    \centering
    \includegraphics[width=\columnwidth]{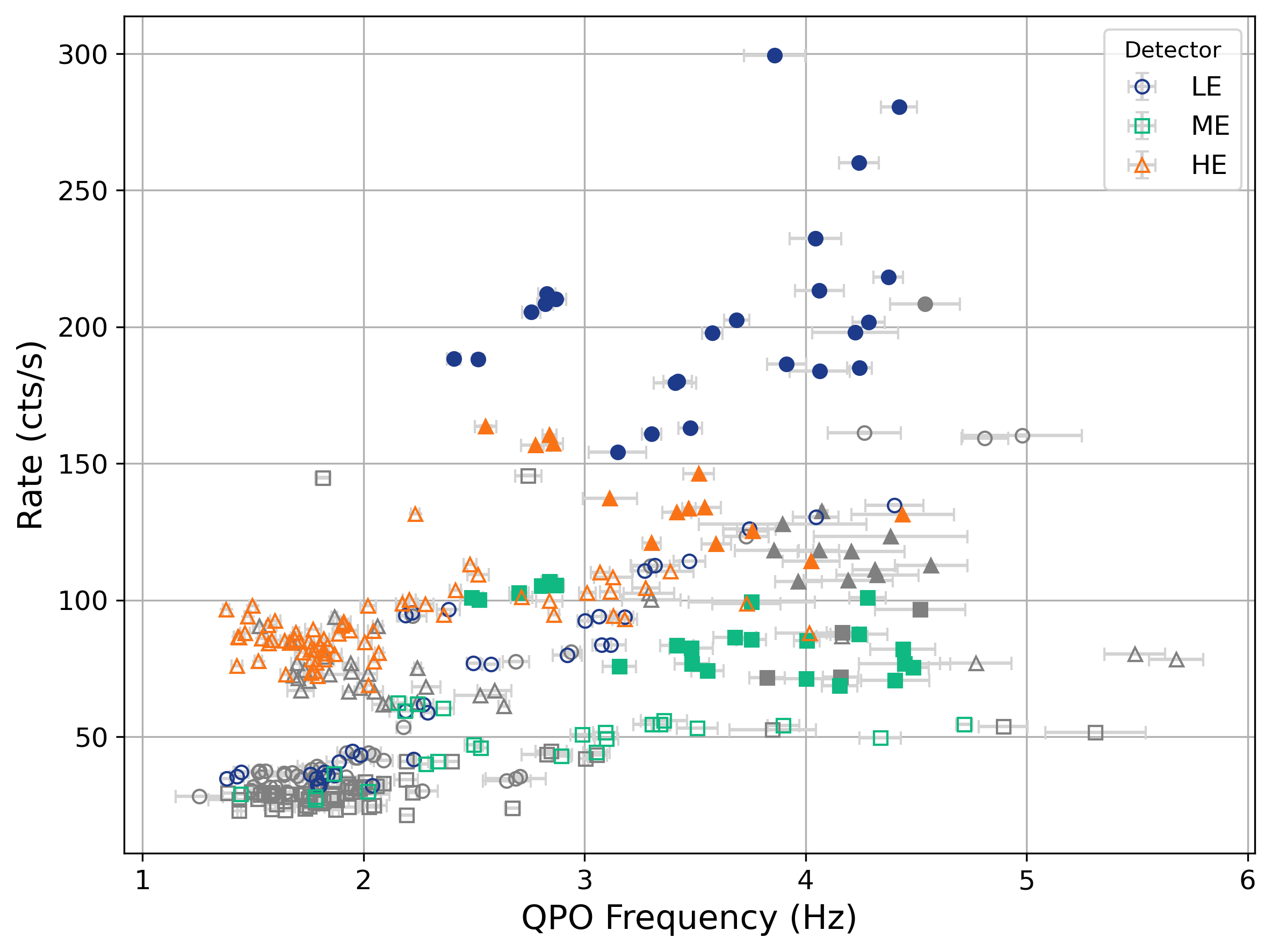}
    \caption{Correlation between count rate and QPO centroid frequency. Data points with significance $\geq$3 are plotted in blue circles (LE), green squares (ME), and orange triangles (HE), while those with significance $<$3 are shown in grey. Filled symbols represent observations during MJD 58200–58230, corresponding to the rising phase of the QPO frequency, while other points are shown as open symbols. All data points are shown with light grey error bars.}
    \label{fig:QPO_LC_relation}
\end{figure}

In Figure~\ref{fig:QPO_LC_relation}, we show the correlation between count rate and QPO frequency. During the QPO frequency increase before MJD~58230 (shown with filled symbols), the LE count rate increases, ME rate remains roughly constant, while HE rate decreases. This trend points to a stronger soft thermal component and weaker high-energy emission, suggesting inward movement of the disk and cooling of the corona. Conversely, during phases of decreasing QPO frequency, the decline in all energy bands indicates suppressed accretion activity and possible recession of the inner disc.

\cite{Heil2011MNRAS.411L..66H} and \cite{Nespoli2003A&A...412..235N} reported correlations between QPO frequency and count rate in the energy range below $\sim$ 15~keV, based on studies of type-C QPO in XTE J1550-564 and type-B QPO in GX 339-4, respectively. Their results are consistent with what we observe during the QPO frequency decrease phase in our analysis. However, the behavior of this correlation at higher energies in other sources remains less explored, and it is still unclear whether similar trends hold in those regimes. In our case, as the QPO frequency increases, the inward movement of the accretion disc enhances thermal emission in the LE band. Meanwhile, the increased soft photon flux may lead to cooling of the Comptonizing corona, thereby reducing the Comptonization efficiency. This results in a decrease in counts in the ME and HE bands, producing a negative correlation between QPO frequency and ME/HE count rates. 

A comparison with MAXI J1535–571 reveals differences. During QPO frequency increases in that source, a similar correlation between QPO frequency and ME/HE count rates is observed. However, during the QPO frequency decrease phase, the ME and HE count rates continue to increase \citep{Chen2022MNRAS.517..182C}, revealing an anti-correlation between them, which is different from GRS 1915+105. This contrast suggests that the decay phase in GRS 1915+105 may have a different evolution process compared to typical outbursts such as MAXI J1535–571. While the corona in MAXI J1535–571 seems to be replenished during its QPO frequency decline phase, the corona in GRS 1915+105 appears to dissipate, possibly transforming into a failed wind gradually that contributes to the observed obscuration.

Recent studies have proposed the existence of two corona components in black hole binary systems \citep{Garcia2021MNRAS.501.3173G,Peirano2023MNRAS.519.1336P,Rawat2023MNRAS.520..113R}. To explain the QPO phase lag observed in GRS 1915+105, it has been suggested that these may arise from two corona regions or two different precessing regions within the accretion flow  \citep{Nobili2000ApJ...538L.137N,Eijnden2016MNRAS.458.3655V}. These two corona are reflected on and are separated by the QPO frequency, where the inner part dominates when QPO frequency exceeds 2~Hz, and the outer part contributes when the frequency is less than 2~Hz \citep{Koljonen2021A&A...647A.173K}. Figure~\ref{fig:QPO_LC_relation} supports the suggestion that the count rate -- QPO frequency relation below and above 2~Hz shows different slope or even different relation. For example, among the open symbols, the slopes of LE and ME data below 2~Hz are flatter then those above 2~Hz, with the difference being more prominent in the LE band. While in the HE band, the slope below 2~Hz becomes even negative.

\subsection{Suspected type-B QPOs}

Type-B QPOs are typically characterized by low rms amplitude ($\sim$ 5\%), narrow peaks (Q $\gtrsim$ 6), and centroid frequencies around 5--6~Hz \citep{Ingram2019NewAR..8501524I}. As mentioned in Section~3, two weak $\sim2$~Hz QPO were detected in the soft state around MJD 58055. To further assess them, we include both in the $\Gamma$--QPO frequency diagram (Figure~\ref{fig:QPO_Gamma_relation}), shown as the two filled red squares. They generally follow the same correlation, although one point lies slightly above the general relation. Since type-B QPOs are commonly associated with soft intermediate state, the suspected type-B QPOs appearance here may suggest a state transition toward the hard state. The MAXI flux shows a sudden decrease immediately after these observations, which would support this interpretation.

As indicated earlier, NICER did not observe GRS 1915+105 on MJD 58055 and MJD 58056, the period when we detected possible type-B QPOs. However, NICER observations resumed on MJD 58057. \cite{Dhaka2024ApJ...974...90D} analyzed this NICER observation along with AstroSat. They reported a prominent broad feature centered at $\sim$ 2~Hz. They employed dynamic power spectra to rule out the possibility of time-varying narrow QPOs and found no such evidence. We also performed wavelet analysis on our data, but due to the weak signal and the intrinsic limitations of wavelet resolution at higher frequencies where the wavelet filter becomes broader in Fourier space, we did not find any prominent signals. Interestingly, We notice that in our observations taken immediately before and after each type-B QPO detection (e.g., P010131000201 and P010131000203 on MJD 58055), the QPO feature in the ME band rapidly evolves into a broad component as mentioned in \cite{Dhaka2024ApJ...974...90D}. This behavior suggests variability in the QPO feature, possibly caused by geometric or dynamical changes in the corona. \cite{Dhaka2024ApJ...974...90D} found that the variability at 2~Hz may affected by both the disk and the corona. However, we find no corresponding QPO signals in the LE band, which is more sensitive to disk emission. This indicates that during these type-B QPOs appear, the disk remains relatively stable while the corona behavior changes.

\subsection{Distinct features of Flare 3 and sub-Hz QPOs}

We note that the obscuration observed in other binary systems such as V404 Cyg and SS 433 are believed to be related to super-Eddington accretion and radio jets \citep{Spencer1979Natur.282..483S,Miller-Jones2019Natur.569..374M}. However, the case of GRS 1915+105 appears to be different. While the observed increase in $T_{\mathrm{in}}$ may suggest enhanced accretion activity, there is no evidence that the source reaches a super-Eddington accretion rate. Furthermore, no significant radio jet actively has been reported during this stage \citep{Motta2021MNRAS.503..152M,Gandhi2025MNRAS.537.1385G}. 

By comparing the spectral-timing properties of the three flares in the obscured state, we find that Flare 3 exhibits a behavior similar to Flare 1. They both show a decrease in $\log \xi$ and an increase in $\Gamma$, while Flare 2 shows the opposite evolution with a higher $\log \xi$ and lower $\Gamma$. However, the presence of the weak $\sim$ 0.2~Hz QPOs in Flare 3 is more similar to Flare 2. Notably, the QPO reported in Flare 2 is only detected up to 18 keV \citep{Kong2024A&A...686A.211K}, which is quite lower than typical type-C QPOs. This may suggest that the obscuration is caused by a Compton-thick outflow \citep{Maiolino1998A&A...338..781M}, potentially associated with the weakened corona in the decay phase. Our data show that Flare 3 is fainter in the LE band but more luminous in the ME band compared to the earlier flares, indicating a more prominent coronal contribution despite enhanced obscuration in soft band.

\cite{Rodriguez2025arXiv250301105R} reported that the accretion disk of GRS 1915+105 became aligned with the line of sight shortly after Flare 3, but returned to its historic inclination in 2024. They estimated the inclination of the ejecta to be $87^{\circ} \pm 3^{\circ}$ at MJD 60217. This temporary alignment may explain the deep X-ray obscuration during that time. Such an extreme viewing angle could increase the blocking effect of the disk, potentially explaining the weak and vague QPO signals observed during Flare 3.

\cite{Gandhi2025MNRAS.537.1385G} report the mid-infrared and radio behavior of GRS 1915+105 before 2024. They found that around MJD 60000, both MIR flux (from NEOWISE) and the radio flux at 15~GHz (from AMI-LA) increased by about an order of magnitude during the X-ray mini-burst prior to Flare 3 (see Figure~\ref{fig:evolve_lc}).The MIR and radio flux behavior of Flare 3 is also similar to that of Flare 1, with both flares being bright in these bands. During Flare 2, although relatively weak radio bursts is still noticed, the MIR flux remains low. If the MIR is contributed by an optically thick wind and its outer nebular fringes as suggested by \cite{Gandhi2025MNRAS.537.1385G}, the presence of the $\sim$ 0.2~Hz QPO in both Flare 2 and Flare 3 indicates that the QPO generation is not related to this wind.

We find a clear difference in hardness ratio between QPOs with frequencies above and below 1~Hz, with lower hardness ratio values typically associated with $>1$~Hz QPOs. This suggests that the contribution from the accretion disk is more significant in these cases, favoring an origin linked to Lense-Thirring precession of the inner hot flow. In contrast, $<1$~Hz QPOs may arise from perturbations in the magnetic field that propagate into a Compton-thick, magnetically driven failed disk wind \citep{Kong2024A&A...686A.211K}. This interpretation is supported by differences in the inferred launching radii of the ionized winds. For the $\sim$0.2~Hz QPOs, \cite{Kong2024A&A...686A.211K} estimated $R_{\rm launch} \lesssim (3$–$9)\times10^4$~km, whereas at the end of the decay phase where $>1$~Hz QPOs are still present, \cite{Zhou2025A&A...694A.104Z} found $R_{\rm launch} \sim 3.5 \times 10^5$~km. Such a more distant wind-launching region is less likely to be magnetically driven and is also less likely to interfere with the generation of $>1$~Hz QPOs near the inner disk. This disparity in wind geometry and coupling thus further supports the hypothesis that high- and low-frequency QPOs may originate from distinct mechanisms.

\section{Conclusion}

In this work, we perform a comprehensive spectral-timing analysis of GRS 1915+105 using Insight-HXMT observations from the soft state to the obscured state. First, We report a previously unstudied QPO frequency rising branch between MJD 58206 and 58230, where the frequency increases from $\sim$2~Hz to 6~Hz, completing the full QPO frequency evolution across a spectral state transition. Second, our detailed timing and spectral analysis of Flare 3 reveals both similarities and differences compared to previous flares, which shows lower ionization and softer spectra. Third, we detect $<1$ Hz QPOs in all three flares, with frequencies spanning $\sim$0.01–0.2 Hz. The QPO signals in Flares 1 and 2 are consistent with earlier reports, while the weak $\sim$0.2 Hz QPO in Flare 3 is newly identified here. Fourth, by comparing QPOs above and below 1~Hz, we find that the differences in hardness ratio and ionized wind launching radius indicate that $>$1~Hz QPOs are likely associated with Lense–Thirring precession of the inner hot flow, while $<$1~Hz QPOs may arise from magnetic perturbations transmitted into a Compton-thick, magnetically driven failed wind. This distinction points to a possible evolution in wind geometry and its coupling with the accretion flow. 

In addition to these four main results, we note tentative evidence for two transient, $\sim$2 Hz type-B QPOs during the soft state (MJD 58055–58056), characterized by low rms amplitudes and high quality factors. However, given their low significance ($\sim 2\sigma$), a firm detection cannot be claimed. Further confirmation with higher signal-to-noise data and independent instruments would be valuable.

These findings not only refine our understanding of the spectral and timing evolution of GRS 1915+105 during its low-flux phase but also highlight the value of the broadband and long-term temporal coverage from Insight-HXMT in revealing key transitional behaviors that were previously missed. Future multi-wavelength observations will be crucial to understand the complex correlations between accretions, outflows, and QPO mechanisms in this source.

\section*{Acknowledgements}

We are grateful to the referee for the useful suggestions to improve the manuscript. This work is supported by the National Natural Science Foundation of China (Grants No. 12435010, 12133007), and the National Key Research and Development Program of China (Grant No. 2022YFA1602301, 2021YFA0718503 and 2023YFA1607901). This work has made use of data from the \textit{Insight}-HXMT mission, a project funded by the China National Space Administration (CNSA) and Chinese Academy of Sciences (CAS).

\section*{Data Availability}

Data used in this work are from the Institute of High Energy Physics, Chinese Academy of Sciences (IHEP-CAS) and have been publicly available for download from the \textit{Insight}-HXMT website http://hxmtweb.ihep.ac.cn/.



\bibliographystyle{mnras}
\bibliography{example} 




\appendix




\bsp	
\label{lastpage}
\end{document}